\documentclass[%
 reprint,
superscriptaddress,
 amsmath,amssymb,
 aps,
pra,
]{revtex4-1}

\usepackage{graphicx}% Include figure files
\usepackage{dcolumn}% Align table columns on decimal point
\usepackage{bm}% bold math
\usepackage[capitalise]{cleveref}
\usepackage{tabularx}
\usepackage{subcaption}

\newcommand\numberthis{\addtocounter{equation}{1}\tag{\theequation}}
\begin{document}
\title{Large Scale Raman Spectrum Calculations in Defective\\2D Materials using Deep Learning}
\author{Olivier Malenfant-Thuot}
\email{olivier.malenfant-thuot@umontreal.ca}
%\affiliation{D\'epartement de Physique, Universit\'e de Montr\'eal}
%\affiliation{Regroupement Qu\'eb\'ecois sur les Mat\'eriaux de Pointe}
\author{Dounia Shaaban Kabakibo}
\email{dounia.shaaban.kabakibo@umontreal.ca}
\affiliation{D\'epartement de Physique et Institut Courtois, Universit\'e de Montr\'eal}
\author{Simon Blackburn}
\author{Bruno Rousseau}
\affiliation{Mila - Québec Artificial Intelligence Institute}
\author{Michel Côté}
\affiliation{D\'epartement de Physique et Institut Courtois, Universit\'e de Montr\'eal}

\begin{abstract}
    We introduce a machine learning prediction workflow to study the impact of defects on the Raman response of 2D materials. By combining the use of machine-learned interatomic potentials, the Raman-active $\Gamma$-weighted density of states method and splitting configurations in independant patches, we are able to reach simulation sizes in the tens of thousands of atoms, with diagonalization now being the main bottleneck of the simulation. We apply the method to two systems, isotopic graphene and defective hexagonal boron nitride, and compare our predicted Raman response to experimental results, with good agreement. Our method opens up many possibilities for future studies of Raman response in solid-state physics.
\end{abstract}

\maketitle

\section{Introduction}
\label{sec:intro}
Raman spectroscopy stands out as a remarkably adaptable and non-invasive technique for the exploration of both molecular and crystalline systems~\cite{rostron2016raman_article4,cialla2019theoretical_article4}. Using the influence of ion motion on electronic states, Raman spectroscopy can provide crucial insights into the examined system, such as the structure, composition, and defects~\cite{orlando2021comprehensive_article4}.

However, it can be challenging to associate features of an experimental spectrum with actual physical insights into the material studied. Therefore, to correctly analyze Raman spectroscopy results, it is useful to be able to simulate theoretical spectrums and compare them to experimental results. Although these predictions can be obtained using \textit{ab initio} methods~\cite{veithen2005nonlinear} for small systems, an in-depth study of defective systems with realistic defect concentrations and varied configurations is essentially beyond reach.

Machine learning models have proven valuable for both analyzing~\cite{luo2022deep,qi2023recent} and simulating Raman spectra~\cite{grisafi2018symmetry_article4,gastegger2021machine_article4,grumet2024delta,malenfant2024efficient}. However, previous simulation workflows have been constrained by their limited capacity to handle only small numbers of atoms, which restricts their ability to study defects. This paper introduces our contribution to this ongoing area of research. Our approach aims to leverage the near-\textit{ab initio} accuracy of machine-learned interatomic potentials (MLIPs) and their significantly faster evaluation times to enable accurate Raman spectrum predictions in large-scale simulation cells.

We applied our method to two 2D systems, namely isotopic graphene and hexagonal boron nitride (hBN) with vacancies. We chose these systems because they are systems of scientific interest~\cite{geim2007rise_article4,armano2019two,molaei2021comprehensive} and because we found high quality experimental results that we could try to reproduce~\cite{carvalho2015probing,li2021defect} using simulations. This paper is organized as follows. Section~\ref{sec:raman_theo} provides the theoretical background  on Raman spectroscopy the basis of our simulation results. Section~\ref{sec:methodology} details the methodology used to obtain the results, beginning with machine-learned potentials, followed by the data generation details and finishing with the large-scale phonon calculations. Results are presented in Section~\ref{sec:results}, where we first discuss isotopic graphene and then defective hBN, comparing our findings with experimental measurements. Finally, Section~\ref{sec:conclusion_article4}, summarizes our findings and explores potential future applications of the methodology.

\section{Raman Spectroscopy Theory}
\label{sec:raman_theo}
When a laser of frequency $\omega_{\textrm{L}}$ is shined on a sample, the Raman peaks for the so-called Stokes and anti-Stokes processes are found at frequencies $\omega_{\textrm{St}}=\omega_{\textrm{L}}-\omega_m$ and $\omega_{\textrm{ASt}}=\omega_{\textrm{L}}+\omega_m$ respectively, where $\omega_m$ is the frequency of the phonon mode $m$. In the Placzek approximation~\cite{placzek1934handbuch}, the Raman scattering efficiency for a Stokes process corresponding to this is~\cite{cardona1982light, veithen2005nonlinear}
\begin{equation} \label{eq:raman_theo}
S^m=C^m\left|\mathbf{e}_\textrm{S} \cdot \boldsymbol{\alpha}^m \cdot \mathbf{e}_\textrm{I}\right|^2,
\end{equation}
with
\begin{equation}
C^m=\frac{\left(\omega_{\textrm{L}}-\omega_m\right)^4}{c^4}\frac{n_m+1}{2\omega_m},
\end{equation}
where $c$ is the speed of light, $\mathbf{e}_\textrm{I}$ and $\mathbf{e}_\textrm{S}$ are the incident and scattered polarizations of the light, $\boldsymbol{\alpha}^m$ is the so-called Raman tensor, and $n_m$ is the phonon occupation factor given by the Bose-Einstein distribution function. The Raman tensor can be computed in terms of the derivative of the electronic susceptibility tensor $\boldsymbol{\chi}$ with respect to atomic positions $X_{n \alpha}$ by 
\begin{equation}
    \alpha_{ij}^m=\sqrt{\Omega_0}\sum_{n=1}^N \sum_{\alpha = 1}^3\frac{\partial \chi_{ij}}{\partial X_{n \alpha}}U_{n \alpha}^m,
\end{equation}
where indices $n$ and $\alpha$ relate to the atom index and cartesian dimension respectively. $U_{n \alpha}^m$ represents the phonon eigendisplacement of mode $m$, derived as solutions to the following generalized eigenvalue problem
\begin{equation}\sum_{n\prime,\alpha\prime}H_{(n\alpha)(n\prime\alpha\prime)}U_{n\prime \alpha\prime}^m = M_n \omega_m^2U_{n \alpha}^m.
\end{equation}
The Hessian matrix $\boldsymbol{H}$ encompasses the interatomic force constants and $M_n$ is the mass of atom $n$. One notes that, because of conservation of momentum, only zone-center phonon modes contribute to first-order Raman scattering. The phonon eigendisplacement modes are chosen such that they satisfy the following orthonormalization relation
\begin{equation}
\sum_{n \alpha}M_n (U_{n \alpha}^m)^*U_{n \alpha}^{m\prime}=\delta_{mm\prime}.\label{eq:orthogonal_relation}
\end{equation}

The theoretical framework governing the Raman process is notably intricate, and performing calculations from first principles for Raman intensities requires knowledge of the dielectric tensor and the system's vibrational modes. In cases where one aims to simulate Raman spectrum at realistic defect distributions, \textit{ab initio} calculations become rapidly demanding. Therefore, we make use of the so-called Raman-active $\Gamma$-weighted density of states (RGDOS) method~\cite{hashemi2019efficient}, based on projection of large supercell (SC) phonons onto those of the primitive cell (PC). When the symmetry of the defective SC is not significantly broken, the modes in the SC that share the same symmetry as the Raman-active modes in the PC can also be considered Raman-active. This allows us to write 
\begin{equation}
\label{eq:projections}
\boldsymbol{\alpha}^{\text{SC},m}\approx \sum _n c_{nm}\boldsymbol{\alpha}^{\text{PC},n}
\end{equation}
where $c_{nm}$ is the projection of the Raman active phonon displacement mode $n$ of the primitive cell  on the phonon displacement mode $m$ of the supercell. By primitive cell we mean the pristine primitive cell, whilst supercell refers to a defective system. The Raman intensities are then obtained using \eqref{eq:raman_theo}, which gives
\begin{equation}
S^{\text{SC},m}\approx \sum c_{nm}^2 S^{\text{PC},n},
\end{equation}
where cross terms have been ignored.
The scheme has already proven successful in various applications, such as transition metal dichalcogenide alloys ($\text{Mo}_x\text{W}_{1-x}\text{S}_2$ and $\text{Zr}\text{S}_x\text{Se}_{1-x}$)~\cite{hashemi2019efficient, oliver2020phonons}, defective $\text{Mo}\text{S}_2$~\cite{kou2020simulating}, and SnS multilayer films~\cite{sutter2021few}. Note that in principle, the projections can only be found by solving a system of linear equations, since the basis is not orthogonal~\cite{berger2023raman}. However, it turns out that the Raman active phonon displacement modes for the primitive cell in graphene and hBN at $\Gamma$, which describe adjacent atoms moving in opposite directions within the plane, are orthogonal, which greatly simplifies the calculations.
In what follows, the Raman intensities we present are the so-called powder averages~\cite{caracas2006theoretical_article4}, given by
\begin{equation}
    S^{\text{(powder)},m}=C^m\left(10G^{(0),m}+5G^{(1),m}+7G^{(2),m}\right),\label{eq:powder}
\end{equation}
where 
\begin{align*}
    G^{(0),m}=&\frac{1}{3}\left( \left| \alpha_{11}^m+\alpha_{22}^m+\alpha_{33}^m \right|^2 \right),\numberthis\label{eq:invariants_article4}\\
    G^{(1),m}=&\frac{1}{2}\left( \left| \alpha_{12}^m-\alpha_{21}^m\right|^2+ \left| \alpha_{13}^m-\alpha_{31}^m\right|^2 +\left| \alpha_{23}^m-\alpha_{32}^m\right|^2 \right),\\
    G^{(2),m}=&\frac{1}{2}\left( \left| \alpha_{12}^m+\alpha_{21}^m\right|^2+ \left| \alpha_{13}^m+\alpha_{31}^m\right|^2 +\left| \alpha_{23}^m+\alpha_{32}^m\right|^2 \right)\\
    +&\frac{1}{3}\left( \left| \alpha_{11}^m-\alpha_{22}^m\right|^2+ \left| \alpha_{11}^m-\alpha_{33}^m\right|^2 +\left| \alpha_{22}^m-\alpha_{33}^m\right|^2 \right).
\end{align*}
This expression correspond to the averaged intensities over incident and scattered light polarizations and all possible crystal orientations.

The Raman tensors for the two $\text{E}_{2g}$ double degenerate Raman-active modes in graphene and hBN are given by~\cite{loudon2001raman}
\begin{equation}
\boldsymbol{\alpha}^{\text{PC},1}=\begin{pmatrix}
0 & d & 0 \\
d & 0 & 0 \\
0 & 0 & 0
\end{pmatrix}
\quad \text{and} \quad
\boldsymbol{\alpha}^{\text{PC},2}=\begin{pmatrix}
d & 0 & 0 \\
0 & -d & 0 \\
0 & 0 & 0
\end{pmatrix},
\end{equation}
which yields $S^{\text{powder},1}=S^{\text{powder},2}=2d^2$. This implies that explicitly computing the Raman intensities of the primitive cell is not required, as all experimental results are reported in arbitrary units, and the only components needed to obtain the Raman intensities in the defective supercell are the $c_{nm}$ coefficients.
\section{Methodology}
\label{sec:methodology}

\subsection{Machine Learned Potentials}
\label{subsec:mlpotentials}

Most modern machine learning frameworks for material sciences are based on atom-centered representations. Individual atomic representations can then be pooled in order to make a global prediction for a specific system. In such a framework, the target quantity, such as total energy, is expressed as
\begin{equation}
    E_T = \sum_{n=1}^N E_n,\label{eq:total_energy}
\end{equation}
where the sum runs overs the $N$ atoms of the configuration. In the current study, we used the SchNet~\cite{schutt2018schnet_article4,schutt2018schnetpack_article4} architecture to represent and fit atomic configurations to energies and forces obtained from density functional theory (DFT). More details on the data are available in Section~\ref{subsec:data}. This architecture is based on a succession of interaction blocks applied to each local atomic environment. The interaction blocs apply gaussian filters that map those local environments to a high dimensional vectorial space. This mapping is invariant under permutation, translation and rotation of the system. After experimentation, and keeping our need for efficient inference in mind, we chose a relatively light set of hyperparameters for our models training, namely a deepness of 2 interaction blocs, an embedding dimensionality of 256 and a radial cut-off of $7\text{\AA}$ for the atomic environments. During training, we used an 80\% train-validation split and batch sizes were set to 8 examples. The learning rate started at a value of $1\times10^{-4}$ and reduced gradually when the validation loss plateaued. Training stopped when the learning rate reached a value of $1\times10^{-6}$.

We should note that many other architectures exist based on representations of local atomic environments. A large part of our Raman workflow is model-agnostic and could be applied to other architectures with an added interface layer. For example, more recent equivariant networks have been shown to perform well for phonon predictions in many materials~\cite{fang2024phonon} and could be of interest in a future study.

Automatic differentiation libraries allow us to obtain gradient values from energy predictions, meaning that we can efficiently calculate forces, as a first-order derivative of the total energy, using
\begin{equation}
    F_{n\alpha} = - \frac{\partial E_T}{\partial X_{n\alpha}}.\label{eq:forces}
\end{equation}
 Similarly, to obtain phonon frequencies and eigenmodes we can compute the Hessian matrix, of dimension $3N\times3N$, using
\begin{equation}
    \label{eq:Hessian}
    H_{(n\alpha) (n\prime \alpha\prime)} = \frac{\partial^2E_T}{\partial X_{n\alpha} \partial X_{n\prime \alpha\prime}} = - \frac{\partial F_{n\alpha}}{\partial X_{n\prime \alpha\prime}},
\end{equation}
which easily transforms into the dynamical matrix $\boldsymbol{D}$ with
\begin{equation}
    D_{(n\alpha) (n\prime \alpha\prime)} = \frac{H_{(n\alpha) (n\prime \alpha\prime)}}{\sqrt{M_n M_{n\prime}}}.\label{eq:dynamical}
\end{equation}
Each eigenvalue and eigenvector pair of $\boldsymbol{D}$ correspond respectively to the squared frequency $\omega^2_m$ and eigenmode $\sqrt{M_n}\boldsymbol{U}^m$ of the $\Gamma$ phonon mode $m$.

\begin{figure*}
    \centering
    \includegraphics[width=\textwidth]{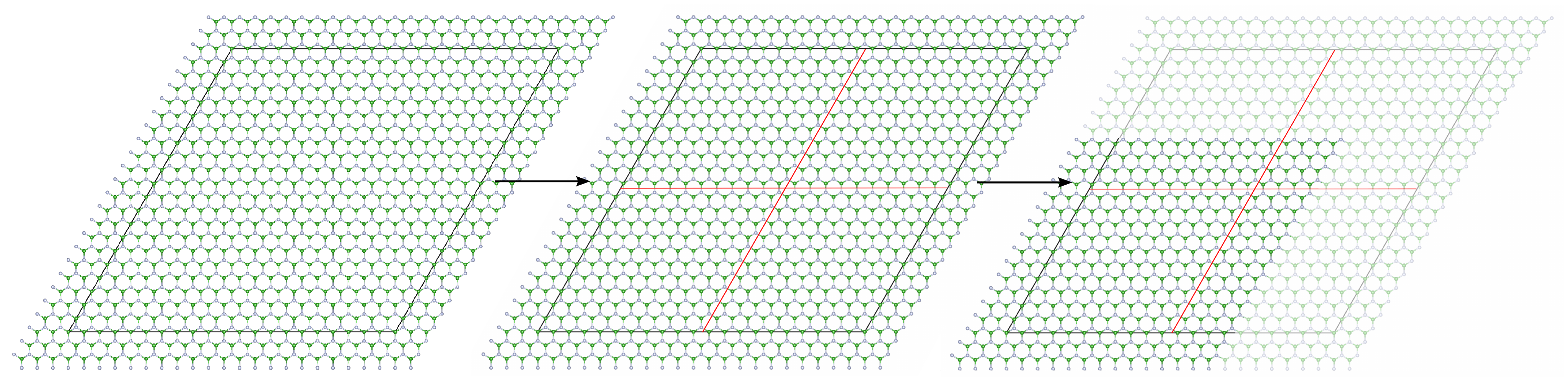}
    \caption{Visual representation a $2\times 2$ patches creation process for an hBN supercell. First step is adding a buffer of new atoms around the supercell. Second step consists of splitting the initial atoms into their own patch. The third step, done separately for each patch, is determining the indices of the main atoms and buffer atoms. Each atom has to be a main atom in a single patch, but can be a buffer atom in zero, one or multiple other patches. The supercell represented here is much smaller than those used in the study.}
    \label{fig:figure_patches}
\end{figure*}

\subsection{Data Generation}
\label{subsec:data}
All our training data was generated using the \textit{ab initio} code ABINIT~\cite{gonze2020abinit_article4, romero2020abinit_article4}. For both graphene and hBN, the training configurations were $7\times7$ supercells containing 98 atoms. We obtained the perturbed positions by running empirical molecular dynamics, with a temperature of 300~K, and keeping one out of every 100 configurations. We then recomputed energy and forces at DFT accuracy using ABINIT. For these calculations, we used the PBE~\cite{Perdew1996} exchange-correlation functional, norm-conserving pseudopotentials~\cite{hamann2013optimized,van2018pseudodojo}, an energy cut-off of 45~Ha and a $4\times4$ unshifted k-points grid. For defective BN configurations, between 3 and 8 random atoms were removed from each configuration, and spin-polarized calculations were needed to account for unpaired electrons in the neighborhood of the vacancies. The self-consistent cycles were stopped once the maximal difference in forces was less than $2\times10^{-5}$ Ha/a.u.

The graphene data set contained a total of 500 perturbed configurations. The hBN data set contained 250 perturbed configurations and the same 250 configurations, but with added vacancies. We completed this data set with a few copies of the relaxation steps of a single vacancy, to have configurations closer to the equilibrium.

\subsection{Phonon Calculations}
\label{subsec:phonon_calc}

The method outlined in Section~\ref{subsec:mlpotentials} gives exact solutions, within the limits of the theory level used, to the phonon eigenmodes problem. However, we aim to scale predictions to very large supercell sizes, which means that two main limitations had to be overcome: GPU memory and exact diagonalization of the dynamical matrix.

First, because of the innate parallelism of GPUs for matrix products, they are much faster than CPUs for training and inference of deep learning models. However, they rely on a finite amount of on-board memory (VRAM) which cannot be extended. During training, this mostly limits the potential size of training batches. During inference, a single configuration is evaluated at a time, but its size will be limited by the amount of memory available. The maximum number of atoms for a given GPU will depend on the model hyperparameters and whether we are computing 1$^{\text{st}}$ order (forces) or 2$^{\text{nd}}$ order (Hessian) derivatives.

In SchNet models, individual atomic representations are updated based on atomic representations of neighboring atoms, which themselves are updated by their neighbors. This means that the effective range ($R_{\textrm{eff}}$) of atomic interactions is actually the number of interaction layers times the cut-off radius, which equates to 14~\AA{} in our case. In other words, the derivative of the energy contribution of atom $n$ with respect to displacements of atom $n\prime$ follows
\begin{equation}
\label{eq:range_buffer}
\begin{aligned}
    \frac{\partial E_n}{\partial X_{n\prime \alpha}} \neq 0& \quad\text{if} \quad |\boldsymbol{X}_n - \boldsymbol{X}_{n\prime}| < R_{\textrm{eff}},\\
    \frac{\partial E_n}{\partial X_{n\prime \alpha}} = 0& \quad \text{if} \quad |\boldsymbol{X}_n - \boldsymbol{X}_{n\prime}| > R_{\textrm{eff}}.
\end{aligned}
\end{equation}

Knowing this, it is possible to split an initial configuration that would be too large to fit in memory into multiple patches that individually satisfy the memory requirements.  Such spatial decomposition techniques are similarly employed in other atomic simulation environments, such as LAMMPS~\cite{thompson2022lammps}. Figure~\ref{fig:figure_patches} visually shows how we achieved it. Each final patch contains two types of atoms: main atoms that were within the boundaries of the patch, and buffer atoms that were outside the boundaries but were distant by less than $R_{\textrm{eff}}$. For patches that share a cell boundary with the initial configuration, the buffer is built by copying atoms from the periodic image. Periodic boundary conditions are dropped in split dimensions, and the patches effectively become large molecules.

Once the configuration is split into patches, a prediction can be made on each of them. We modified the SchNet models so that they could account for two sets of indices. In the representation layers, all atoms of the patches are treated equally, but in the prediction layers, only the main atoms of the patch contribute to the final value. In this framework, Equation~\ref{eq:total_energy} becomes
\begin{equation}
    E_T = \sum_{p=1}^P \left(\sum_{n=1}^{N_p} E_n \right),
\end{equation}
where $P$ is the total number of patches and $N_p$ is the number of main atoms in patch $p$. Equations~\ref{eq:forces} and \ref{eq:Hessian} can be computed within individual patches and mapped back to the atoms of the initial configuration. Equation~\ref{eq:range_buffer} confirms us that the derivatives obtained for atom $n$ are the same we would have obtained in the initial configuration, as long as it is a main atom of the patch from which we obtain the value.

The second limitation in phonon calculations is the diagonalization of the dynamical matrix obtained by Equation~\ref{eq:dynamical}. While this step can be negligible for small systems, it becomes very resource intensive when considering thousands of atoms. Linear algebra routines for exact diagonalization tend to scale as $N^3$ in time and $N^2$ in memory, with multiple copies of the matrix stored. Therefore, for our graphene simulations, we used the implicitly restarted Lanczos method~\cite{lehoucq1998arpack}, as implemented in SciPy~\cite{2020SciPy-NMeth}, to converge the highest frequency modes of the dynamical matrix. It is an iterative method that relies on a matrix-vector product operation to construct an orthogonal basis piece by piece and project the matrix onto that space. That product does not need to be computed explicitly; it can be replaced by a proxy function giving the same result. The effect of the dynamical matrix ($\boldsymbol{D}$) on a displacement vector ($\boldsymbol{V}$) can be expressed as
\begin{align*}
        (\boldsymbol{D} \cdot \boldsymbol{V})_{n\alpha} &= \sum_{n\prime=1}^{N} \sum_{\alpha\prime=1}^{3} \frac{\partial F_{n\alpha}}{\partial X_{n\prime\alpha\prime}} \cdot V_{n\prime\alpha\prime},\numberthis\label{eq:matvec}\\
        &= \frac{\boldsymbol{F}(\boldsymbol{X}+ \xi \boldsymbol{V}) - \boldsymbol{F}(\boldsymbol{X}-\xi \boldsymbol{V})}{2\xi} + O(\xi^2).
\end{align*}
This allows us to replace the matrix-vector product with only two force evaluations in the direction of $\boldsymbol{V}$. We chose $\xi$ so that the average atomic displacement had a norm of 0.01~\AA{}.

Using this property, we were able to compute a fraction of the system's phonon frequencies and eigenmodes without ever explicitly computing the dynamical matrix. The higher this fraction, the longer the Lanczos algorithm takes to converge. For this reason, we limited the diagonalization calculations to the amount of eigenmodes necessary to obtain 99\% of the total projection on the two active Raman modes, which means that the squared sum of the $c_{nm}$ coefficients in Equation~\ref{eq:projections} equals at least 0.99. In the case of a pristine supercell, all the projection is contained within the two corresponding eigenmodes, but with the inclusion of more and more defects, a larger range of eigenmodes is necessary to reach this threshold, up to 16\% of the complete eigenbasis for isotopic graphene. Using these projections and Equations~\ref{eq:projections}, \ref{eq:powder} and \ref{eq:invariants_article4}, we computed the final Raman intensities in the supercell with defects. We aimed to use the same technique for the hBN part of our study, but since vacancies are a much more disrupting defect than isotopes, we would have to compute more than a large majority of the eigenvectors to reach the same coverage of 99\%. Therefore, in that case, we limited ourselves to simulation cell sizes that could be diagonalized exactly in a reasonable amount of time.

The codes used for to run the simulations are available online~\cite{codes1,codes2}.

\section{Results}
\label{sec:results}
We applied the methodology from Section~\ref{subsec:phonon_calc} to investigate the impact of defects on two different 2D systems: isotopic graphene and hBN with vacancies. The two following sections will present the results obtained for each of these cases.

\begin{figure*}
    \centering
    \begin{subfigure}[t]{0.45\textwidth}
        \includegraphics[width=\textwidth]{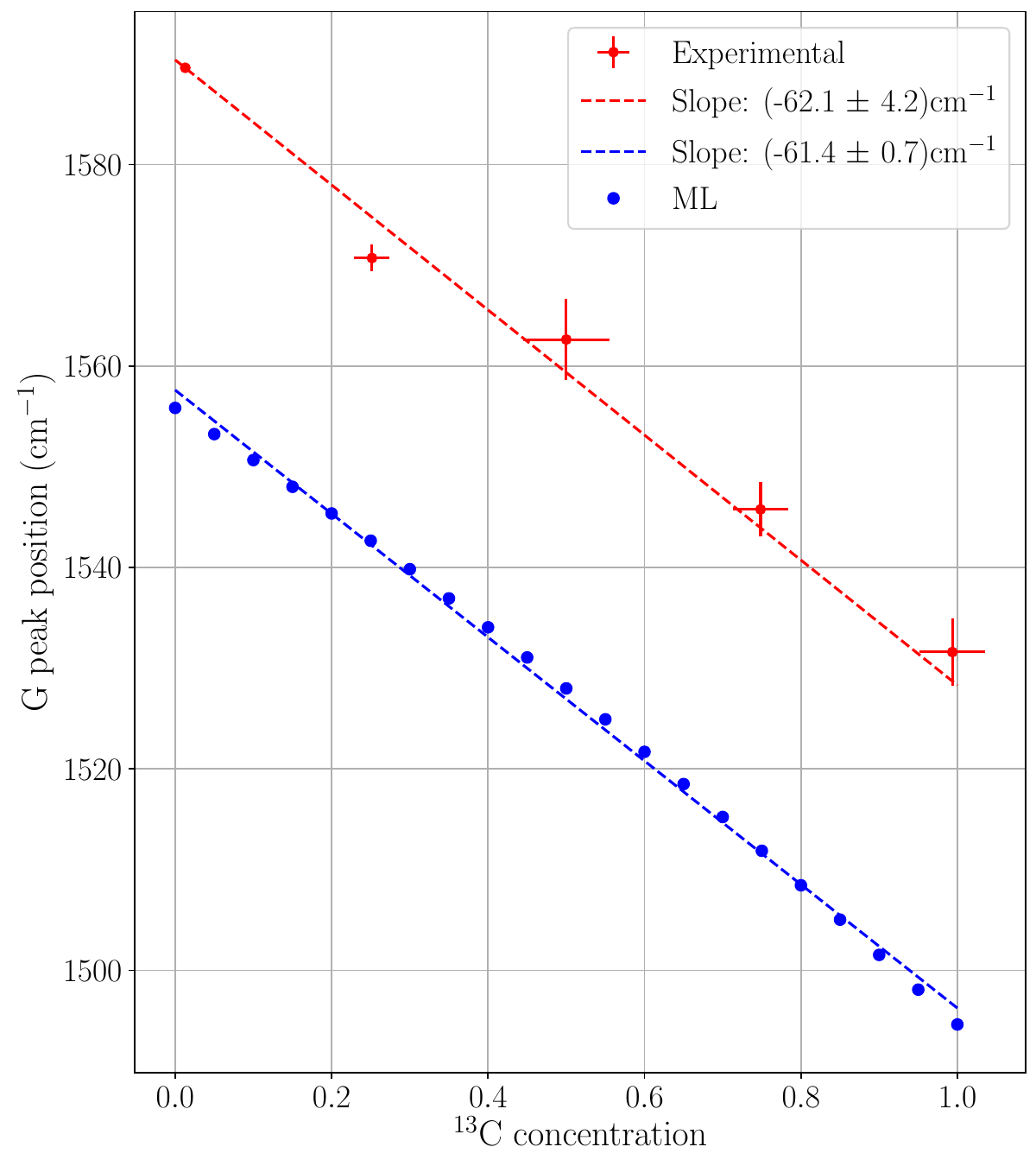}
        \caption{Experimental and simulated center frequency of the G peak}
    \end{subfigure}
    \hfill
    \begin{subfigure}[t]{0.45\textwidth}
        \includegraphics[width=\textwidth]{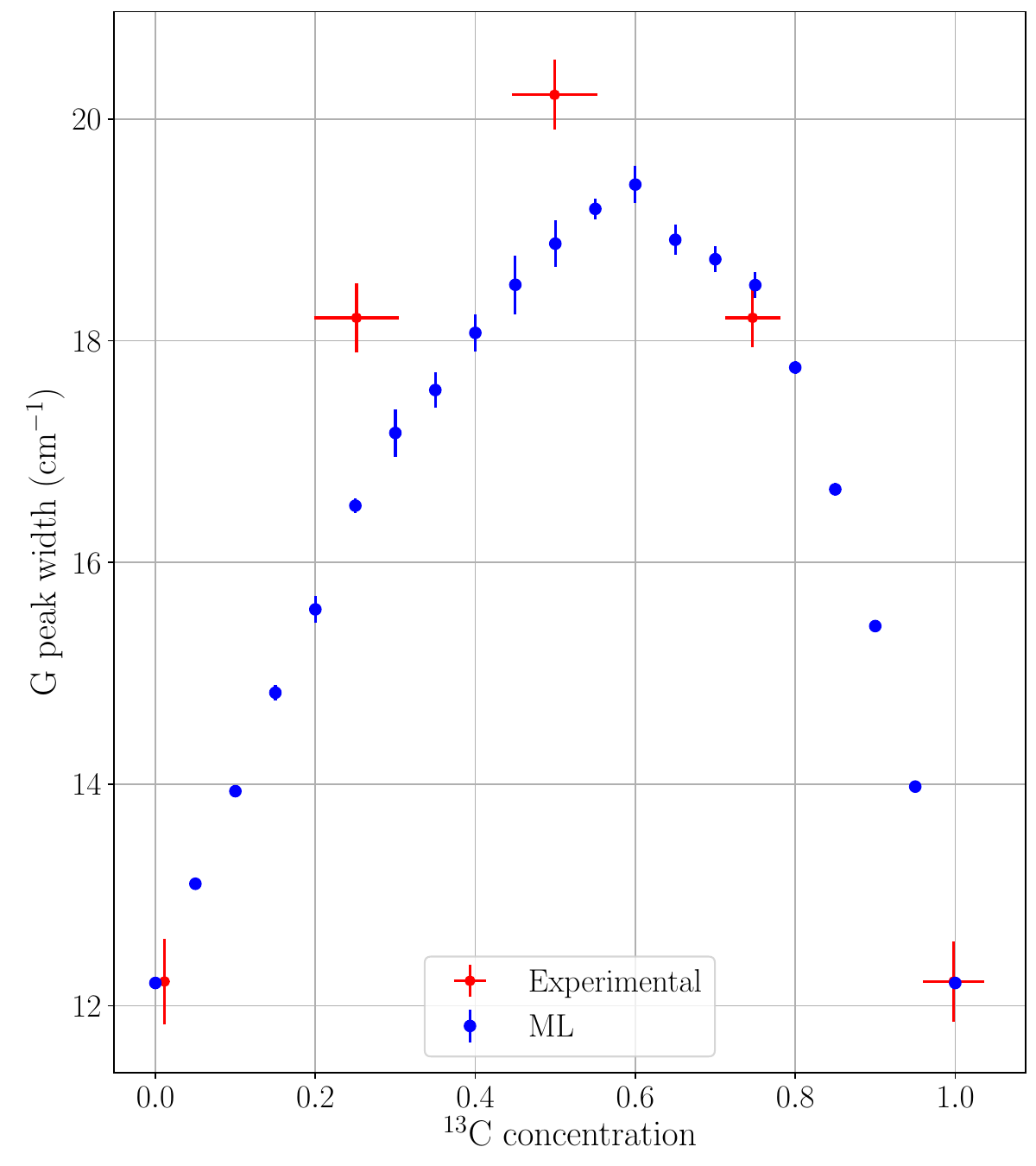}
        \caption{Experimental and simulated FWHM of the G peak}
    \end{subfigure}
    \caption{Evolution of the G peak of isotopic graphene with varying concentrations of $^{12}$C and $^{13}$C. The error bars in the simulated results show the standard deviation between different configurations of same concentration. Experimental results are from ~\cite{carvalho2015probing}.}
    \label{fig:isotopes_graphene}
\end{figure*}

\subsection{Isotopic Graphene}
\label{subsec:isotopes}
Raman spectroscopy has been used for some time as a way to probe the structural integrity of graphene and assess the effectiveness of various doping and treatment methods~\cite{canccado2011quantifying,wu2018raman}. While some empirical models have been devised to explain the experimental observations~\cite{canccado2017disentangling}, we were able to obtain a near \textit{ab initio} simulation of a large-scale observable phenomena.

The excellent experimental work by Carvalho et al.~\cite{carvalho2015probing} served as an anchor point for this simulation. In that study, different concentrations of pure $^{12}$C and $^{13}$C were used to grow graphene samples with a controlled amount of each isotope. Raman spectroscopy was then used to measure the impact of the relative concentrations on the obtained spectrums. All experimental values presented in Section~\ref{subsec:isotopes} originate from this study. We studied the evolution of the center frequency and full width at half-maximum (FWHM) of the main G peak of graphene and compared with the experimental values.

The iterative Lanczos method described in Section~\ref{subsec:phonon_calc} was used with $115\times115$ graphene supercells containing 26,450 atoms. We explored the relative concentrations of $^{12}$C from 100\% to 0\% with steps of 5\%, filling the remaining carbons with $^{13}$C. For each concentration value, except for 0\% and 100\% where a single configuration exists, we generated 3 random distributions of isotopes and applied our Raman prediction workflow to them individually, using a $3\times3$ patch grid to fit forces calculations in GPU memory. Each Lanczos process took around 5 days to complete on a single GPU and needed at most 25~GB of RAM. In comparison, the exact diagonalization process using Scipy would take about 10 $\textrm{cpu} \cdot \textrm{days}$, which can be parallelized to some extent, and necessitate around 110~GB of RAM. The computational gain is not that significant at first glance, but the Lanczos method scales as $N^2$ in timing, while the exact diagonalization scales as $N^3$, where $N$ is the number of atoms. For future work, where we aim to reach even higher simulation sizes, this advantage will be significant. Moreover, the GPUs used for this calculations were NVidia V100 with 16~GB memories; using compute units with larger memories would allow for larger patches and greater efficiency.

Once each Raman spectrum was obtained, a Lorentzian function was fitted on the G peak, and its parameters represented the values we used as center and width for the physical peak. We then aggregated those parameters to produce the results visible in Figure~\ref{fig:isotopes_graphene}.

At first, we can observe a shift between the distributions of the experimental and simulated center frequencies. This shift of around $30~\textrm{cm}^{-1}$ or 4~meV is acceptable within DFT accuracy. More importantly, the observed slopes in both cases match well and represent the physical phenomenon that we are trying to capture with this simulation. We know that increasing the masses of atoms will decrease their vibrational frequencies, but here we were able to compute precisely that effect with sufficient statistics. The results proved very stable between simulations, to the point that the bars representing the standard deviations would have been smaller than the markers used. Interestingly, we observe a slight concavity in the simulated results, while a simple $\sqrt(k/m)$ fit would give a convex function. Some anharmonic effects on the carbon structure probably play a role in that observation. We note that there are not enough experimental data to infer the curvature of the experimental values.

Next, we can study the FWHM of the simulated G peaks. Experimentally, the measured width of the G peak has two contributions, an intrinsic one related to phonon lifetimes and a second one related to disorder~\cite{bonini2007phonon}. Our method can evaluate that second contribution only, therefore we added an initial smearing width to the active modes to match the observed experimental width at 0\% and 100\% concentrations. That initial intrinsic width of around 12~cm$^{-1}$ is supported by other works~\cite{bonini2007phonon,liu2019intrinsic}. We kept the smearing parameter fixed for all the other simulations and the G peak got larger accordingly with increasing disorder in the structure. We can observe some variance in the FWHM simulated results, especially in the middle concentrations, where the number of possible configurations of isotopes is higher, but the results are still stable enough to easily analyze trends. Other than these configurations, variance is mostly negligible, meaning the statistics reached within a single supercell are already sufficient to capture most of the contributing factors. Our simulated results came very close to the experimental values, although we obtained a slight asymmetry in the curve that did not seem to be observed experimentally.

Overall, these results show that the method can capture modifications to Raman intensities from increased disorder in a material. The logical next step was the study of bigger perturbations.

\subsection{Defective BN}
\label{subsec:bn}

\begin{figure}
    \centering
    \includegraphics[width=\linewidth]{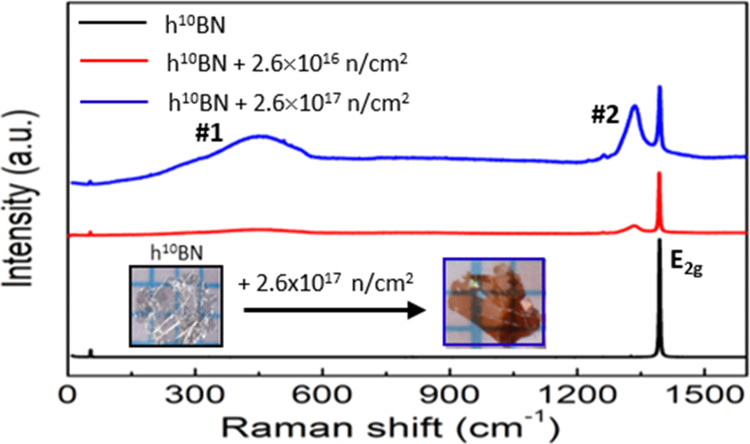}
    \caption{Raman spectrums of hBN with increasing levels of B vacancies induced by neutron irradiation. Reprinted with permission from~\cite{li2021defect}. Copyright 2021 American Chemical Society.}
    \label{fig:experimental_BN_raman}
\end{figure}

A recent experiment was conducted by Li \textit{et al.}~\cite{li2021defect} on multilayer hBN. Their experimental setup utilized neutron irradiation to selectively induce Boron vacancies in the sample, with the vacancy concentration expected to be proportional to the exposure time. The impact of treatment on Raman spectrums, with different exposure times, is visible in Figure~\ref{fig:experimental_BN_raman}, reproduced from their report. The black line is the Raman response of the untreated sample, the red line is obtained after a short exposure to the treatment and the blue line is obtained after a much longer exposure. Two new peaks, labeled \#1 and \#2, are clearly visible at high exposure, while the main $E_{\textrm{2g}}$ is not perturbed. This represents an interesting system for our method as hBN has a significant electronic gap, meaning the appearance of new peaks is not driven by resonant Raman processes, like in graphene. The weakly active mode at very low frequency is an interlayer vibration mode in hBN and is not present in our single layer simulations.

We simulated this system in $95\times95\times1$ supercells, containing 18,050 atoms before adding vacancies. To obtain both peaks, the full eigenspectrum was needed, so we explicitly computed the Hessian matrix and diagonalized it in a second separate step. We studied varying values of vacancies concentration, from 0\% to 3.5\%, with steps of 0.5\%. After removing randomly selected B atoms , we relaxed the obtained configuration so that the atoms rest at the minimum of the potentiel energy surface learned by the model. This step proved harder than initially expected, especially at concentrations above 2\%. This is probably due to the trained potential having difficulties precisely modeling the energy near equilibrium when vacancies are close by, because that type of configuration is not included in its training set. Therefore, for the higher concentrations, we had to generate multiple configurations and select the few ones that could relax the best, which biases our random sampling towards configurations with more even distributions of vacancies. That also limited our search to a maximum concentration of 3.5\%. Forces were calculated on a $2\times2\times1$ patches grid for the relaxations.

\begin{figure}
    \centering
    \includegraphics[width=\linewidth]{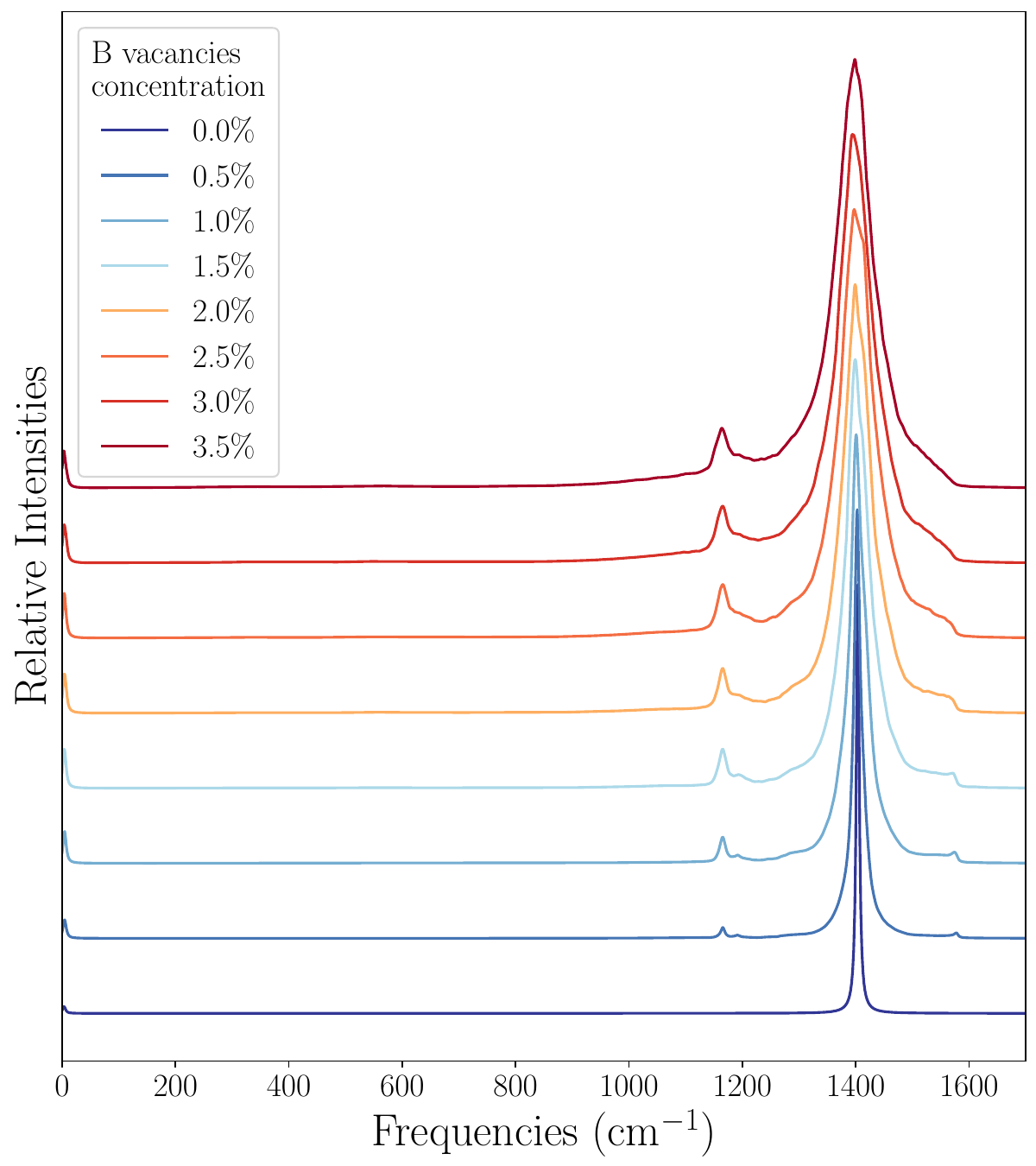}
    \caption{Evolution of the simulated Raman spectrums of defective hBN with different concentrations of B vacancies.}
    \label{fig:B_raman_spectrums}
\end{figure}

Once relaxed, the supercells were split on a $4\times4\times1$ patches grid for the Hessian calculations. Those were completed in only 2 hours on a single GPU. The diagonalizations took about 8 cpus$\cdot$hours and 35~GB of RAM. The obtained eigenmodes were projected on the eigendisplacements corresponding to the active modes in pristine hBN. To respect dimensionality between the defective and pristine modes, the indices of the removed atoms were initially saved and the corresponding rows in the pristine mode were dropped. We repeated the process 4 times per concentration and averaged results to have more significant statistics, but we did not find much variation between samples. The Raman intensities obtained are visible in Figure~\ref{fig:B_raman_spectrums}.

\begin{figure}
    \centering
    \includegraphics[width=\linewidth]{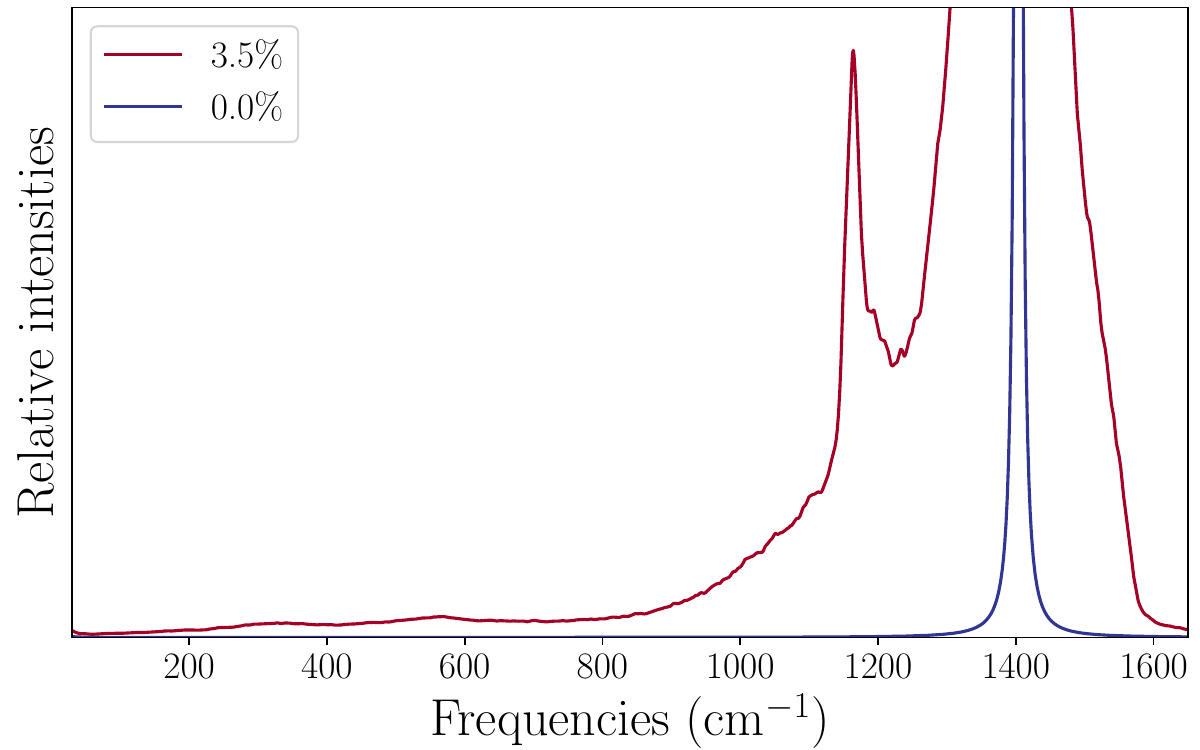}
    \caption{Comparison of the Raman intensities at the highest simulated concentration of B vacancies with the pristine system.}
    \label{fig:B_comparison}
\end{figure}

The main features observable in these results are the appearance of a new Raman peak around 1200~cm$^{-1}$ and a broadening of the original peak. Although this does not agree perfectly with the experimental results of Figure~\ref{fig:experimental_BN_raman}, there is qualitative agreement with the appearance of a new peak (\#2) at slightly lower frequency than the main one. To better observe the changes induced by defects, Figure~\ref{fig:B_comparison} only plots the pristine and 3.5\% concentration intensities and zooms in on the lowest part of the spectrum. In this representation, it is easier to see that the main peak, although wider, has not shifted significantly; the maximum intensity appears at 1404~cm$^{-1}$ without defects and at 1398~cm$^{-1}$ with them. In addition, we can observe the appearance of a wide Raman peak between 200 and 700~cm$^{-1}$, which seems to correspond to mode \#1. In general, our results look much more like the red curve in Figure~\ref{fig:experimental_BN_raman} than the other ones. With a fluence 10 times higher to obtain the blue curve, we can roughly extrapolate and predict that this spectrum would correspond to a higher vacancy concentration in the 20\% or 30\% range.

We can also observe the appearance of a small sharp peak at very low frequencies in the plotted intensities of Figure~\ref{fig:B_raman_spectrums}. This is not due to a physical process but to numerical errors in the model and the method. Acoustic phonon bands in graphene and hBN have a particular quadratic dispersion that is known to be difficult to capture with empirical potentials~\cite{adamyan2010phonons,koukaras2015phonon} and we have observed many times that trained potentials have difficulties predicting very low frequency phonon modes even when they have great performance everywhere else. In addition, this peak could also originate from the treatment of vacancies in the pristine mode, since nullifying the displacement of the removed atoms makes the mode gain a small amount of momentum, meaning it is no longer a pure $q=0$ mode. Improving the method to get rid of this numerical noise would be worthwhile, but we do not expect it to change the overall results.

\begin{figure}
    \centering
    \includegraphics[width=\linewidth]{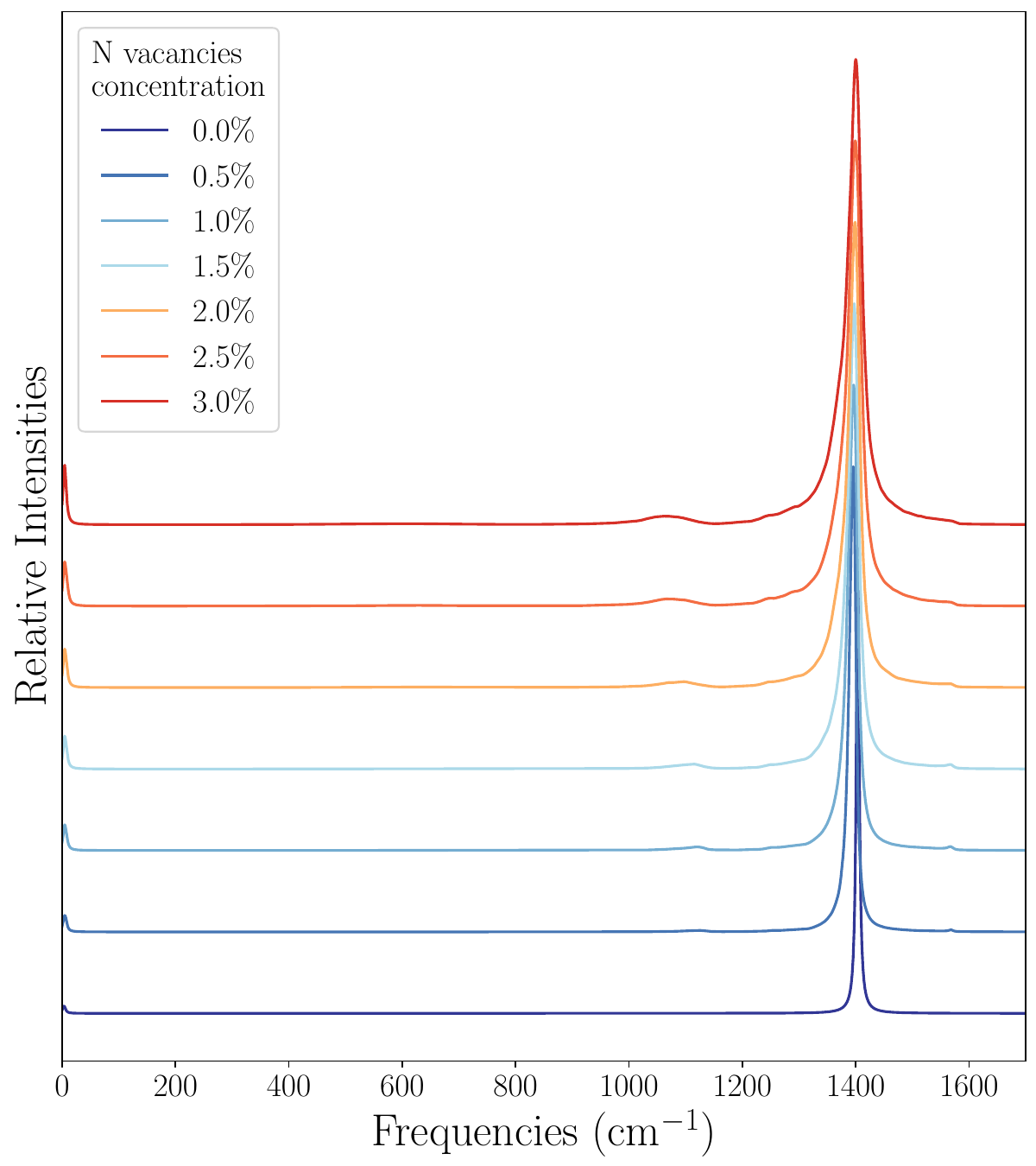}
    \caption{Evolution of the simulated Raman spectrums of defective hBN with different concentrations of N vacancies.}
    \label{fig:N_raman_spectrums}
\end{figure}

With the method tested on B vacancies, we can use it on N vacancies, even though we do not have experimental results in that case. The same methodology was used as previously. The only notable difference is that the N vacancies were slightly harder to relax than the B vacancies, so we had to limit configurations to a 3\% concentration or less. The results are available in Figure~\ref{fig:N_raman_spectrums}.

Interestingly, we do not observe the same Raman features in this case. While the main peak somewhat broadens, it does not do so nearly as much as in the B vacancies case. Furthermore, there is a small augmentation of intensities below 1200~cm$^{-1}$, but it is much less strong and wider than the relatively sharp peak previously observed. Overall, the N vacancies seem to have much less of an impact on the Raman intensities of hBN than the B vacancies, an observation that could be important in future analysis of experimental results.

\subsection{Future Work}

While the results we presented in Section~\ref{sec:results} already provide us with a new perspective on Raman response simulations, we can anticipate further enhancements in the method and its broader applications.

First, improving models performance for low frequencies phonon predictions and for relaxation with higher defects concentration would both improve the precision of the obtained intensities and allow for study of highly defective systems. These challenges might be surmountable by iterating on the data generation strategies and neural network architectures. Second, a more rigorous treatment of vacancies or additional atoms in the pristine modes could help reduce numerical noise and improve the reliability of the results. Third, as a bigger goal, being able to incorporate resonant Raman response into the simulations would allow us to study problems of even greater interest to the scientific community, like the apparition of the D and 2D peaks in defective graphene.

On the topic of simulation scale, the patches method allows a very efficient computation of the Hessian matrix, with nearly linear scaling when increasing the grid size accordingly with the simulation cell. The issue now lies with the diagonalization of the dynamical matrix. Exploring other ways to perform this operation on matrices that get more and more sparse as they grow could allow to reach even higher sizes and simulate even more realistic systems.

\section{Conclusion}
\label{sec:conclusion_article4}
In this work, we introduced a machine learning prediction workflow for Raman spectrum simulations in large-scale supercells that can include realistic defect configurations in a statistically significant way. The patches method decouples the simulation size from the available GPU memory and allows to efficiently calculate forces and Hessian matrices past the size limit of the original potentials. We applied the method to predict the Raman response of isotopic graphene and got excellent agreement on the evolution of the G peak frequency and width with respect to the concentration of isotopes compared to experimental measurements. We also predicted the Raman response of defective hBN with increasing vacancies concentration and calculated the appearance of a new peak at lower frequency than the main one, a phenomenon that was observed experimentally. Overall, this method allows for much larger simulation sizes than previously possible and should be usable to study a large range of physical systems.

\section*{Acknowledgments}
\label{sec:ack}
We thank Pr. Hannu-Pekka Komsa for kindly answering our interrogations. This research was financially supported by the Natural Sciences and Engineering Research Council of Canada (NSERC), under the Discovery Grants program Grant No. RGPIN-2016-06666 (M.C.), and the National Research Council of Canada (NRC) under the Artificial Intelligence for Design Project. Computations were made on the supercomputers Beluga and Narval managed by Calcul Québec and the Digital Research Alliance of Canada. The operation of these supercomputers is funded by the Canada Foundation for Innovation, the Ministère de la Science, de l’Économie et de l’Innovation du Québec, and the Fonds de recherche du Québec – Nature et technologies. D.S.K.  acknowledges support by the NSERC Postgraduate Scholarships – Doctoral program and the FRQ Nature and technologies Sector – Doctoral Research Scholarship Program. M.C. is a member of the Regroupement québécois sur les matériaux de pointe (RQMP).

\bibliography{refs}

\begin{thebibliography}{44}%
\makeatletter
\providecommand \@ifxundefined [1]{%
 \@ifx{#1\undefined}
}%
\providecommand \@ifnum [1]{%
 \ifnum #1\expandafter \@firstoftwo
 \else \expandafter \@secondoftwo
 \fi
}%
\providecommand \@ifx [1]{%
 \ifx #1\expandafter \@firstoftwo
 \else \expandafter \@secondoftwo
 \fi
}%
\providecommand \natexlab [1]{#1}%
\providecommand \enquote  [1]{``#1''}%
\providecommand \bibnamefont  [1]{#1}%
\providecommand \bibfnamefont [1]{#1}%
\providecommand \citenamefont [1]{#1}%
\providecommand \href@noop [0]{\@secondoftwo}%
\providecommand \href [0]{\begingroup \@sanitize@url \@href}%
\providecommand \@href[1]{\@@startlink{#1}\@@href}%
\providecommand \@@href[1]{\endgroup#1\@@endlink}%
\providecommand \@sanitize@url [0]{\catcode `\\12\catcode `\$12\catcode `\&12\catcode `\#12\catcode `\^12\catcode `\_12\catcode `\%12\relax}%
\providecommand \@@startlink[1]{}%
\providecommand \@@endlink[0]{}%
\providecommand \url  [0]{\begingroup\@sanitize@url \@url }%
\providecommand \@url [1]{\endgroup\@href {#1}{\urlprefix }}%
\providecommand \urlprefix  [0]{URL }%
\providecommand \Eprint [0]{\href }%
\providecommand \doibase [0]{http://dx.doi.org/}%
\providecommand \selectlanguage [0]{\@gobble}%
\providecommand \bibinfo  [0]{\@secondoftwo}%
\providecommand \bibfield  [0]{\@secondoftwo}%
\providecommand \translation [1]{[#1]}%
\providecommand \BibitemOpen [0]{}%
\providecommand \bibitemStop [0]{}%
\providecommand \bibitemNoStop [0]{.\EOS\space}%
\providecommand \EOS [0]{\spacefactor3000\relax}%
\providecommand \BibitemShut  [1]{\csname bibitem#1\endcsname}%
\let\auto@bib@innerbib\@empty
%</preamble>
\bibitem [{\citenamefont {Rostron}\ \emph {et~al.}(2016)\citenamefont {Rostron}, \citenamefont {Gaber},\ and\ \citenamefont {Gaber}}]{rostron2016raman_article4}%
  \BibitemOpen
  \bibfield  {author} {\bibinfo {author} {\bibfnamefont {P.}~\bibnamefont {Rostron}}, \bibinfo {author} {\bibfnamefont {S.}~\bibnamefont {Gaber}}, \ and\ \bibinfo {author} {\bibfnamefont {D.}~\bibnamefont {Gaber}},\ }\href@noop {} {\bibfield  {journal} {\bibinfo  {journal} {Laser}\ }\textbf {\bibinfo {volume} {21}},\ \bibinfo {pages} {24} (\bibinfo {year} {2016})}\BibitemShut {NoStop}%
\bibitem [{\citenamefont {Cialla-May}\ \emph {et~al.}(2019)\citenamefont {Cialla-May}, \citenamefont {Schmitt},\ and\ \citenamefont {Popp}}]{cialla2019theoretical_article4}%
  \BibitemOpen
  \bibfield  {author} {\bibinfo {author} {\bibfnamefont {D.}~\bibnamefont {Cialla-May}}, \bibinfo {author} {\bibfnamefont {M.}~\bibnamefont {Schmitt}}, \ and\ \bibinfo {author} {\bibfnamefont {J.}~\bibnamefont {Popp}},\ }\href@noop {} {\bibfield  {journal} {\bibinfo  {journal} {Physical Sciences Reviews}\ }\textbf {\bibinfo {volume} {4}} (\bibinfo {year} {2019})}\BibitemShut {NoStop}%
\bibitem [{\citenamefont {Orlando}\ \emph {et~al.}(2021)\citenamefont {Orlando}, \citenamefont {Franceschini}, \citenamefont {Muscas}, \citenamefont {Pidkova}, \citenamefont {Bartoli}, \citenamefont {Rovere},\ and\ \citenamefont {Tagliaferro}}]{orlando2021comprehensive_article4}%
  \BibitemOpen
  \bibfield  {author} {\bibinfo {author} {\bibfnamefont {A.}~\bibnamefont {Orlando}}, \bibinfo {author} {\bibfnamefont {F.}~\bibnamefont {Franceschini}}, \bibinfo {author} {\bibfnamefont {C.}~\bibnamefont {Muscas}}, \bibinfo {author} {\bibfnamefont {S.}~\bibnamefont {Pidkova}}, \bibinfo {author} {\bibfnamefont {M.}~\bibnamefont {Bartoli}}, \bibinfo {author} {\bibfnamefont {M.}~\bibnamefont {Rovere}}, \ and\ \bibinfo {author} {\bibfnamefont {A.}~\bibnamefont {Tagliaferro}},\ }\href@noop {} {\bibfield  {journal} {\bibinfo  {journal} {Chemosensors}\ }\textbf {\bibinfo {volume} {9}},\ \bibinfo {pages} {262} (\bibinfo {year} {2021})}\BibitemShut {NoStop}%
\bibitem [{\citenamefont {Veithen}\ \emph {et~al.}(2005)\citenamefont {Veithen}, \citenamefont {Gonze},\ and\ \citenamefont {Ghosez}}]{veithen2005nonlinear}%
  \BibitemOpen
  \bibfield  {author} {\bibinfo {author} {\bibfnamefont {M.}~\bibnamefont {Veithen}}, \bibinfo {author} {\bibfnamefont {X.}~\bibnamefont {Gonze}}, \ and\ \bibinfo {author} {\bibfnamefont {P.}~\bibnamefont {Ghosez}},\ }\href@noop {} {\bibfield  {journal} {\bibinfo  {journal} {Physical Review B—Condensed Matter and Materials Physics}\ }\textbf {\bibinfo {volume} {71}},\ \bibinfo {pages} {125107} (\bibinfo {year} {2005})}\BibitemShut {NoStop}%
\bibitem [{\citenamefont {Luo}\ \emph {et~al.}(2022)\citenamefont {Luo}, \citenamefont {Popp},\ and\ \citenamefont {Bocklitz}}]{luo2022deep}%
  \BibitemOpen
  \bibfield  {author} {\bibinfo {author} {\bibfnamefont {R.}~\bibnamefont {Luo}}, \bibinfo {author} {\bibfnamefont {J.}~\bibnamefont {Popp}}, \ and\ \bibinfo {author} {\bibfnamefont {T.}~\bibnamefont {Bocklitz}},\ }\href@noop {} {\bibfield  {journal} {\bibinfo  {journal} {Analytica}\ }\textbf {\bibinfo {volume} {3}},\ \bibinfo {pages} {287} (\bibinfo {year} {2022})}\BibitemShut {NoStop}%
\bibitem [{\citenamefont {Qi}\ \emph {et~al.}(2023)\citenamefont {Qi}, \citenamefont {Hu}, \citenamefont {Jiang}, \citenamefont {Wu}, \citenamefont {Zheng}, \citenamefont {Chen}, \citenamefont {Liang}, \citenamefont {Sadi}, \citenamefont {Zhang},\ and\ \citenamefont {Chen}}]{qi2023recent}%
  \BibitemOpen
  \bibfield  {author} {\bibinfo {author} {\bibfnamefont {Y.}~\bibnamefont {Qi}}, \bibinfo {author} {\bibfnamefont {D.}~\bibnamefont {Hu}}, \bibinfo {author} {\bibfnamefont {Y.}~\bibnamefont {Jiang}}, \bibinfo {author} {\bibfnamefont {Z.}~\bibnamefont {Wu}}, \bibinfo {author} {\bibfnamefont {M.}~\bibnamefont {Zheng}}, \bibinfo {author} {\bibfnamefont {E.~X.}\ \bibnamefont {Chen}}, \bibinfo {author} {\bibfnamefont {Y.}~\bibnamefont {Liang}}, \bibinfo {author} {\bibfnamefont {M.~A.}\ \bibnamefont {Sadi}}, \bibinfo {author} {\bibfnamefont {K.}~\bibnamefont {Zhang}}, \ and\ \bibinfo {author} {\bibfnamefont {Y.~P.}\ \bibnamefont {Chen}},\ }\href@noop {} {\bibfield  {journal} {\bibinfo  {journal} {Advanced Optical Materials}\ }\textbf {\bibinfo {volume} {11}},\ \bibinfo {pages} {2203104} (\bibinfo {year} {2023})}\BibitemShut {NoStop}%
\bibitem [{\citenamefont {Grisafi}\ \emph {et~al.}(2018)\citenamefont {Grisafi}, \citenamefont {Wilkins}, \citenamefont {Cs{\'a}nyi},\ and\ \citenamefont {Ceriotti}}]{grisafi2018symmetry_article4}%
  \BibitemOpen
  \bibfield  {author} {\bibinfo {author} {\bibfnamefont {A.}~\bibnamefont {Grisafi}}, \bibinfo {author} {\bibfnamefont {D.~M.}\ \bibnamefont {Wilkins}}, \bibinfo {author} {\bibfnamefont {G.}~\bibnamefont {Cs{\'a}nyi}}, \ and\ \bibinfo {author} {\bibfnamefont {M.}~\bibnamefont {Ceriotti}},\ }\href@noop {} {\bibfield  {journal} {\bibinfo  {journal} {Physical Review Letters}\ }\textbf {\bibinfo {volume} {120}},\ \bibinfo {pages} {036002} (\bibinfo {year} {2018})}\BibitemShut {NoStop}%
\bibitem [{\citenamefont {Gastegger}\ \emph {et~al.}(2021)\citenamefont {Gastegger}, \citenamefont {Sch{\"u}tt},\ and\ \citenamefont {M{\"u}ller}}]{gastegger2021machine_article4}%
  \BibitemOpen
  \bibfield  {author} {\bibinfo {author} {\bibfnamefont {M.}~\bibnamefont {Gastegger}}, \bibinfo {author} {\bibfnamefont {K.~T.}\ \bibnamefont {Sch{\"u}tt}}, \ and\ \bibinfo {author} {\bibfnamefont {K.-R.}\ \bibnamefont {M{\"u}ller}},\ }\href@noop {} {\bibfield  {journal} {\bibinfo  {journal} {Chemical Science}\ }\textbf {\bibinfo {volume} {12}},\ \bibinfo {pages} {11473} (\bibinfo {year} {2021})}\BibitemShut {NoStop}%
\bibitem [{\citenamefont {Grumet}\ \emph {et~al.}(2024)\citenamefont {Grumet}, \citenamefont {von Scarpatetti}, \citenamefont {Bučko},\ and\ \citenamefont {Egger}}]{grumet2024delta}%
  \BibitemOpen
  \bibfield  {author} {\bibinfo {author} {\bibfnamefont {M.}~\bibnamefont {Grumet}}, \bibinfo {author} {\bibfnamefont {C.}~\bibnamefont {von Scarpatetti}}, \bibinfo {author} {\bibfnamefont {T.}~\bibnamefont {Bučko}}, \ and\ \bibinfo {author} {\bibfnamefont {D.~A.}\ \bibnamefont {Egger}},\ }\href@noop {} {\bibfield  {journal} {\bibinfo  {journal} {The Journal of Physical Chemistry C}\ }\textbf {\bibinfo {volume} {128}},\ \bibinfo {pages} {6464} (\bibinfo {year} {2024})}\BibitemShut {NoStop}%
\bibitem [{\citenamefont {Malenfant-Thuot}\ \emph {et~al.}(2024{\natexlab{a}})\citenamefont {Malenfant-Thuot}, \citenamefont {Ryczko}, \citenamefont {Tamblyn},\ and\ \citenamefont {Cote}}]{malenfant2024efficient}%
  \BibitemOpen
  \bibfield  {author} {\bibinfo {author} {\bibfnamefont {O.}~\bibnamefont {Malenfant-Thuot}}, \bibinfo {author} {\bibfnamefont {K.}~\bibnamefont {Ryczko}}, \bibinfo {author} {\bibfnamefont {I.}~\bibnamefont {Tamblyn}}, \ and\ \bibinfo {author} {\bibfnamefont {M.}~\bibnamefont {Cote}},\ }\href {http://iopscience.iop.org/article/10.1088/1361-648X/ad64a2} {\bibfield  {journal} {\bibinfo  {journal} {Journal of Physics: Condensed Matter}\ } (\bibinfo {year} {2024}{\natexlab{a}})}\BibitemShut {NoStop}%
\bibitem [{\citenamefont {Geim}\ and\ \citenamefont {Novoselov}(2007)}]{geim2007rise_article4}%
  \BibitemOpen
  \bibfield  {author} {\bibinfo {author} {\bibfnamefont {A.~K.}\ \bibnamefont {Geim}}\ and\ \bibinfo {author} {\bibfnamefont {K.~S.}\ \bibnamefont {Novoselov}},\ }\href@noop {} {\bibfield  {journal} {\bibinfo  {journal} {Nature materials}\ }\textbf {\bibinfo {volume} {6}},\ \bibinfo {pages} {183} (\bibinfo {year} {2007})}\BibitemShut {NoStop}%
\bibitem [{\citenamefont {Armano}\ and\ \citenamefont {Agnello}(2019)}]{armano2019two}%
  \BibitemOpen
  \bibfield  {author} {\bibinfo {author} {\bibfnamefont {A.}~\bibnamefont {Armano}}\ and\ \bibinfo {author} {\bibfnamefont {S.}~\bibnamefont {Agnello}},\ }\href@noop {} {\bibfield  {journal} {\bibinfo  {journal} {C}\ }\textbf {\bibinfo {volume} {5}},\ \bibinfo {pages} {67} (\bibinfo {year} {2019})}\BibitemShut {NoStop}%
\bibitem [{\citenamefont {Molaei}\ \emph {et~al.}(2021)\citenamefont {Molaei}, \citenamefont {Younas},\ and\ \citenamefont {Rezakazemi}}]{molaei2021comprehensive}%
  \BibitemOpen
  \bibfield  {author} {\bibinfo {author} {\bibfnamefont {M.~J.}\ \bibnamefont {Molaei}}, \bibinfo {author} {\bibfnamefont {M.}~\bibnamefont {Younas}}, \ and\ \bibinfo {author} {\bibfnamefont {M.}~\bibnamefont {Rezakazemi}},\ }\href@noop {} {\bibfield  {journal} {\bibinfo  {journal} {ACS Applied Electronic Materials}\ }\textbf {\bibinfo {volume} {3}},\ \bibinfo {pages} {5165} (\bibinfo {year} {2021})}\BibitemShut {NoStop}%
\bibitem [{\citenamefont {Carvalho}\ \emph {et~al.}(2015)\citenamefont {Carvalho}, \citenamefont {Hao}, \citenamefont {Righi}, \citenamefont {Rodriguez-Nieva}, \citenamefont {Colombo}, \citenamefont {Ruoff}, \citenamefont {Pimenta},\ and\ \citenamefont {Fantini}}]{carvalho2015probing}%
  \BibitemOpen
  \bibfield  {author} {\bibinfo {author} {\bibfnamefont {B.~R.}\ \bibnamefont {Carvalho}}, \bibinfo {author} {\bibfnamefont {Y.}~\bibnamefont {Hao}}, \bibinfo {author} {\bibfnamefont {A.}~\bibnamefont {Righi}}, \bibinfo {author} {\bibfnamefont {J.~F.}\ \bibnamefont {Rodriguez-Nieva}}, \bibinfo {author} {\bibfnamefont {L.}~\bibnamefont {Colombo}}, \bibinfo {author} {\bibfnamefont {R.~S.}\ \bibnamefont {Ruoff}}, \bibinfo {author} {\bibfnamefont {M.~A.}\ \bibnamefont {Pimenta}}, \ and\ \bibinfo {author} {\bibfnamefont {C.}~\bibnamefont {Fantini}},\ }\href@noop {} {\bibfield  {journal} {\bibinfo  {journal} {Physical Review B}\ }\textbf {\bibinfo {volume} {92}},\ \bibinfo {pages} {125406} (\bibinfo {year} {2015})}\BibitemShut {NoStop}%
\bibitem [{\citenamefont {Li}\ \emph {et~al.}(2021)\citenamefont {Li}, \citenamefont {Glaser}, \citenamefont {Elias}, \citenamefont {Ye}, \citenamefont {Evans}, \citenamefont {Xue}, \citenamefont {Liu}, \citenamefont {Cassabois}, \citenamefont {Gil}, \citenamefont {Valvin} \emph {et~al.}}]{li2021defect}%
  \BibitemOpen
  \bibfield  {author} {\bibinfo {author} {\bibfnamefont {J.}~\bibnamefont {Li}}, \bibinfo {author} {\bibfnamefont {E.~R.}\ \bibnamefont {Glaser}}, \bibinfo {author} {\bibfnamefont {C.}~\bibnamefont {Elias}}, \bibinfo {author} {\bibfnamefont {G.}~\bibnamefont {Ye}}, \bibinfo {author} {\bibfnamefont {D.}~\bibnamefont {Evans}}, \bibinfo {author} {\bibfnamefont {L.}~\bibnamefont {Xue}}, \bibinfo {author} {\bibfnamefont {S.}~\bibnamefont {Liu}}, \bibinfo {author} {\bibfnamefont {G.}~\bibnamefont {Cassabois}}, \bibinfo {author} {\bibfnamefont {B.}~\bibnamefont {Gil}}, \bibinfo {author} {\bibfnamefont {P.}~\bibnamefont {Valvin}},  \emph {et~al.},\ }\href@noop {} {\bibfield  {journal} {\bibinfo  {journal} {Chemistry of Materials}\ }\textbf {\bibinfo {volume} {33}},\ \bibinfo {pages} {9231} (\bibinfo {year} {2021})}\BibitemShut {NoStop}%
\bibitem [{\citenamefont {Placzek}(1934)}]{placzek1934handbuch}%
  \BibitemOpen
  \bibfield  {author} {\bibinfo {author} {\bibfnamefont {G.}~\bibnamefont {Placzek}},\ }\href@noop {} {\bibfield  {journal} {\bibinfo  {journal} {Ed. G. Marx, Akademische Verlagsgesellschaft, Leipzig}\ } (\bibinfo {year} {1934})}\BibitemShut {NoStop}%
\bibitem [{\citenamefont {Cardona}\ and\ \citenamefont {G{\"u}ntherodt}(1982)}]{cardona1982light}%
  \BibitemOpen
  \bibfield  {author} {\bibinfo {author} {\bibfnamefont {M.}~\bibnamefont {Cardona}}\ and\ \bibinfo {author} {\bibfnamefont {G.}~\bibnamefont {G{\"u}ntherodt}},\ }\href@noop {} {\emph {\bibinfo {title} {Light Scattering in Solids {II}}}}\ (\bibinfo  {publisher} {Springer},\ \bibinfo {year} {1982})\BibitemShut {NoStop}%
\bibitem [{\citenamefont {Hashemi}\ \emph {et~al.}(2019)\citenamefont {Hashemi}, \citenamefont {Krasheninnikov}, \citenamefont {Puska},\ and\ \citenamefont {Komsa}}]{hashemi2019efficient}%
  \BibitemOpen
  \bibfield  {author} {\bibinfo {author} {\bibfnamefont {A.}~\bibnamefont {Hashemi}}, \bibinfo {author} {\bibfnamefont {A.~V.}\ \bibnamefont {Krasheninnikov}}, \bibinfo {author} {\bibfnamefont {M.}~\bibnamefont {Puska}}, \ and\ \bibinfo {author} {\bibfnamefont {H.-P.}\ \bibnamefont {Komsa}},\ }\href@noop {} {\bibfield  {journal} {\bibinfo  {journal} {Physical Review Materials}\ }\textbf {\bibinfo {volume} {3}},\ \bibinfo {pages} {023806} (\bibinfo {year} {2019})}\BibitemShut {NoStop}%
\bibitem [{\citenamefont {Oliver}\ \emph {et~al.}(2020)\citenamefont {Oliver}, \citenamefont {Fox}, \citenamefont {Hashemi}, \citenamefont {Singh}, \citenamefont {Cavalero}, \citenamefont {Yee}, \citenamefont {Snyder}, \citenamefont {Jaramillo}, \citenamefont {Komsa},\ and\ \citenamefont {Vora}}]{oliver2020phonons}%
  \BibitemOpen
  \bibfield  {author} {\bibinfo {author} {\bibfnamefont {S.~M.}\ \bibnamefont {Oliver}}, \bibinfo {author} {\bibfnamefont {J.~J.}\ \bibnamefont {Fox}}, \bibinfo {author} {\bibfnamefont {A.}~\bibnamefont {Hashemi}}, \bibinfo {author} {\bibfnamefont {A.}~\bibnamefont {Singh}}, \bibinfo {author} {\bibfnamefont {R.~L.}\ \bibnamefont {Cavalero}}, \bibinfo {author} {\bibfnamefont {S.}~\bibnamefont {Yee}}, \bibinfo {author} {\bibfnamefont {D.~W.}\ \bibnamefont {Snyder}}, \bibinfo {author} {\bibfnamefont {R.}~\bibnamefont {Jaramillo}}, \bibinfo {author} {\bibfnamefont {H.-P.}\ \bibnamefont {Komsa}}, \ and\ \bibinfo {author} {\bibfnamefont {P.~M.}\ \bibnamefont {Vora}},\ }\href@noop {} {\bibfield  {journal} {\bibinfo  {journal} {Journal of Materials Chemistry C}\ }\textbf {\bibinfo {volume} {8}},\ \bibinfo {pages} {5732} (\bibinfo {year} {2020})}\BibitemShut {NoStop}%
\bibitem [{\citenamefont {Kou}\ \emph {et~al.}(2020)\citenamefont {Kou}, \citenamefont {Hashemi}, \citenamefont {Puska}, \citenamefont {Krasheninnikov},\ and\ \citenamefont {Komsa}}]{kou2020simulating}%
  \BibitemOpen
  \bibfield  {author} {\bibinfo {author} {\bibfnamefont {Z.}~\bibnamefont {Kou}}, \bibinfo {author} {\bibfnamefont {A.}~\bibnamefont {Hashemi}}, \bibinfo {author} {\bibfnamefont {M.~J.}\ \bibnamefont {Puska}}, \bibinfo {author} {\bibfnamefont {A.~V.}\ \bibnamefont {Krasheninnikov}}, \ and\ \bibinfo {author} {\bibfnamefont {H.-P.}\ \bibnamefont {Komsa}},\ }\href@noop {} {\bibfield  {journal} {\bibinfo  {journal} {npj Computational Materials}\ }\textbf {\bibinfo {volume} {6}},\ \bibinfo {pages} {59} (\bibinfo {year} {2020})}\BibitemShut {NoStop}%
\bibitem [{\citenamefont {Sutter}\ \emph {et~al.}(2021)\citenamefont {Sutter}, \citenamefont {Komsa}, \citenamefont {Lu}, \citenamefont {Gruverman},\ and\ \citenamefont {Sutter}}]{sutter2021few}%
  \BibitemOpen
  \bibfield  {author} {\bibinfo {author} {\bibfnamefont {P.}~\bibnamefont {Sutter}}, \bibinfo {author} {\bibfnamefont {H.}~\bibnamefont {Komsa}}, \bibinfo {author} {\bibfnamefont {H.}~\bibnamefont {Lu}}, \bibinfo {author} {\bibfnamefont {A.}~\bibnamefont {Gruverman}}, \ and\ \bibinfo {author} {\bibfnamefont {E.}~\bibnamefont {Sutter}},\ }\href@noop {} {\bibfield  {journal} {\bibinfo  {journal} {Nano Today}\ }\textbf {\bibinfo {volume} {37}},\ \bibinfo {pages} {101082} (\bibinfo {year} {2021})}\BibitemShut {NoStop}%
\bibitem [{\citenamefont {Berger}\ \emph {et~al.}(2023)\citenamefont {Berger}, \citenamefont {Lv},\ and\ \citenamefont {Komsa}}]{berger2023raman}%
  \BibitemOpen
  \bibfield  {author} {\bibinfo {author} {\bibfnamefont {E.}~\bibnamefont {Berger}}, \bibinfo {author} {\bibfnamefont {Z.-P.}\ \bibnamefont {Lv}}, \ and\ \bibinfo {author} {\bibfnamefont {H.-P.}\ \bibnamefont {Komsa}},\ }\href@noop {} {\bibfield  {journal} {\bibinfo  {journal} {Journal of Materials Chemistry C}\ }\textbf {\bibinfo {volume} {11}},\ \bibinfo {pages} {1311} (\bibinfo {year} {2023})}\BibitemShut {NoStop}%
\bibitem [{\citenamefont {Caracas}\ and\ \citenamefont {Cohen}(2006)}]{caracas2006theoretical_article4}%
  \BibitemOpen
  \bibfield  {author} {\bibinfo {author} {\bibfnamefont {R.}~\bibnamefont {Caracas}}\ and\ \bibinfo {author} {\bibfnamefont {R.~E.}\ \bibnamefont {Cohen}},\ }\href@noop {} {\bibfield  {journal} {\bibinfo  {journal} {Geophysical Research Letters}\ }\textbf {\bibinfo {volume} {33}} (\bibinfo {year} {2006})}\BibitemShut {NoStop}%
\bibitem [{\citenamefont {Loudon}(2001)}]{loudon2001raman}%
  \BibitemOpen
  \bibfield  {author} {\bibinfo {author} {\bibfnamefont {R.}~\bibnamefont {Loudon}},\ }\href@noop {} {\bibfield  {journal} {\bibinfo  {journal} {Advances in Physics}\ }\textbf {\bibinfo {volume} {50}},\ \bibinfo {pages} {813} (\bibinfo {year} {2001})}\BibitemShut {NoStop}%
\bibitem [{\citenamefont {et~al.}(2018)}]{schutt2018schnet_article4}%
  \BibitemOpen
  \bibfield  {author} {\bibinfo {author} {\bibfnamefont {K.~S.}\ \bibnamefont {et~al.}},\ }\href@noop {} {\bibfield  {journal} {\bibinfo  {journal} {The {Journal} of {Chemical} {Physics}}\ }\textbf {\bibinfo {volume} {148}},\ \bibinfo {pages} {241722} (\bibinfo {year} {2018})}\BibitemShut {NoStop}%
\bibitem [{\citenamefont {Schütt}\ \emph {et~al.}(2018)\citenamefont {Schütt}, \citenamefont {Kessel}, \citenamefont {Gastegger}, \citenamefont {Nicoli}, \citenamefont {Tkatchenko},\ and\ \citenamefont {Müller}}]{schutt2018schnetpack_article4}%
  \BibitemOpen
  \bibfield  {author} {\bibinfo {author} {\bibfnamefont {K.}~\bibnamefont {Schütt}}, \bibinfo {author} {\bibfnamefont {P.}~\bibnamefont {Kessel}}, \bibinfo {author} {\bibfnamefont {M.}~\bibnamefont {Gastegger}}, \bibinfo {author} {\bibfnamefont {K.}~\bibnamefont {Nicoli}}, \bibinfo {author} {\bibfnamefont {A.}~\bibnamefont {Tkatchenko}}, \ and\ \bibinfo {author} {\bibfnamefont {K.-R.}\ \bibnamefont {Müller}},\ }\href@noop {} {\bibfield  {journal} {\bibinfo  {journal} {Journal of chemical theory and computation}\ }\textbf {\bibinfo {volume} {15}},\ \bibinfo {pages} {448} (\bibinfo {year} {2018})}\BibitemShut {NoStop}%
\bibitem [{\citenamefont {Fang}\ \emph {et~al.}(2024)\citenamefont {Fang}, \citenamefont {Geiger}, \citenamefont {Checkelsky},\ and\ \citenamefont {Smidt}}]{fang2024phonon}%
  \BibitemOpen
  \bibfield  {author} {\bibinfo {author} {\bibfnamefont {S.}~\bibnamefont {Fang}}, \bibinfo {author} {\bibfnamefont {M.}~\bibnamefont {Geiger}}, \bibinfo {author} {\bibfnamefont {J.~G.}\ \bibnamefont {Checkelsky}}, \ and\ \bibinfo {author} {\bibfnamefont {T.}~\bibnamefont {Smidt}},\ }\href@noop {} {\bibfield  {journal} {\bibinfo  {journal} {arXiv preprint arXiv:2403.11347}\ } (\bibinfo {year} {2024})}\BibitemShut {NoStop}%
\bibitem [{\citenamefont {Gonze}\ \emph {et~al.}(2020)\citenamefont {Gonze}, \citenamefont {Amadon}, \citenamefont {Antonius}, \citenamefont {Arnardi}, \citenamefont {Baguet}, \citenamefont {Beuken}, \citenamefont {Bieder}, \citenamefont {Bottin}, \citenamefont {Bouchet}, \citenamefont {Bousquet} \emph {et~al.}}]{gonze2020abinit_article4}%
  \BibitemOpen
  \bibfield  {author} {\bibinfo {author} {\bibfnamefont {X.}~\bibnamefont {Gonze}}, \bibinfo {author} {\bibfnamefont {B.}~\bibnamefont {Amadon}}, \bibinfo {author} {\bibfnamefont {G.}~\bibnamefont {Antonius}}, \bibinfo {author} {\bibfnamefont {F.}~\bibnamefont {Arnardi}}, \bibinfo {author} {\bibfnamefont {L.}~\bibnamefont {Baguet}}, \bibinfo {author} {\bibfnamefont {J.-M.}\ \bibnamefont {Beuken}}, \bibinfo {author} {\bibfnamefont {J.}~\bibnamefont {Bieder}}, \bibinfo {author} {\bibfnamefont {F.}~\bibnamefont {Bottin}}, \bibinfo {author} {\bibfnamefont {J.}~\bibnamefont {Bouchet}}, \bibinfo {author} {\bibfnamefont {E.}~\bibnamefont {Bousquet}},  \emph {et~al.},\ }\href@noop {} {\bibfield  {journal} {\bibinfo  {journal} {Computer Physics Communications}\ }\textbf {\bibinfo {volume} {248}},\ \bibinfo {pages} {107042} (\bibinfo {year} {2020})}\BibitemShut {NoStop}%
\bibitem [{\citenamefont {Romero}\ \emph {et~al.}(2020)\citenamefont {Romero}, \citenamefont {Allan}, \citenamefont {Amadon}, \citenamefont {Antonius}, \citenamefont {Applencourt}, \citenamefont {Baguet}, \citenamefont {Bieder}, \citenamefont {Bottin}, \citenamefont {Bouchet}, \citenamefont {Bousquet} \emph {et~al.}}]{romero2020abinit_article4}%
  \BibitemOpen
  \bibfield  {author} {\bibinfo {author} {\bibfnamefont {A.~H.}\ \bibnamefont {Romero}}, \bibinfo {author} {\bibfnamefont {D.~C.}\ \bibnamefont {Allan}}, \bibinfo {author} {\bibfnamefont {B.}~\bibnamefont {Amadon}}, \bibinfo {author} {\bibfnamefont {G.}~\bibnamefont {Antonius}}, \bibinfo {author} {\bibfnamefont {T.}~\bibnamefont {Applencourt}}, \bibinfo {author} {\bibfnamefont {L.}~\bibnamefont {Baguet}}, \bibinfo {author} {\bibfnamefont {J.}~\bibnamefont {Bieder}}, \bibinfo {author} {\bibfnamefont {F.}~\bibnamefont {Bottin}}, \bibinfo {author} {\bibfnamefont {J.}~\bibnamefont {Bouchet}}, \bibinfo {author} {\bibfnamefont {E.}~\bibnamefont {Bousquet}},  \emph {et~al.},\ }\href@noop {} {\bibfield  {journal} {\bibinfo  {journal} {The Journal of Chemical Physics}\ }\textbf {\bibinfo {volume} {152}},\ \bibinfo {pages} {124102} (\bibinfo {year} {2020})}\BibitemShut {NoStop}%
\bibitem [{\citenamefont {Perdew}\ \emph {et~al.}(1996)\citenamefont {Perdew}, \citenamefont {Burke},\ and\ \citenamefont {Ernzerhof}}]{Perdew1996}%
  \BibitemOpen
  \bibfield  {author} {\bibinfo {author} {\bibfnamefont {J.~P.}\ \bibnamefont {Perdew}}, \bibinfo {author} {\bibfnamefont {K.}~\bibnamefont {Burke}}, \ and\ \bibinfo {author} {\bibfnamefont {M.}~\bibnamefont {Ernzerhof}},\ }\href {\doibase 10.1103/physrevlett.77.3865} {\bibfield  {journal} {\bibinfo  {journal} {Phys. Rev. Lett.}\ }\textbf {\bibinfo {volume} {77}},\ \bibinfo {pages} {3865} (\bibinfo {year} {1996})}\BibitemShut {NoStop}%
\bibitem [{\citenamefont {Hamann}(2013)}]{hamann2013optimized}%
  \BibitemOpen
  \bibfield  {author} {\bibinfo {author} {\bibfnamefont {D.}~\bibnamefont {Hamann}},\ }\href@noop {} {\bibfield  {journal} {\bibinfo  {journal} {Physical Review B}\ }\textbf {\bibinfo {volume} {88}},\ \bibinfo {pages} {085117} (\bibinfo {year} {2013})}\BibitemShut {NoStop}%
\bibitem [{\citenamefont {Van~Setten}\ \emph {et~al.}(2018)\citenamefont {Van~Setten}, \citenamefont {Giantomassi}, \citenamefont {Bousquet}, \citenamefont {Verstraete}, \citenamefont {Hamann}, \citenamefont {Gonze},\ and\ \citenamefont {Rignanese}}]{van2018pseudodojo}%
  \BibitemOpen
  \bibfield  {author} {\bibinfo {author} {\bibfnamefont {M.~J.}\ \bibnamefont {Van~Setten}}, \bibinfo {author} {\bibfnamefont {M.}~\bibnamefont {Giantomassi}}, \bibinfo {author} {\bibfnamefont {E.}~\bibnamefont {Bousquet}}, \bibinfo {author} {\bibfnamefont {M.~J.}\ \bibnamefont {Verstraete}}, \bibinfo {author} {\bibfnamefont {D.~R.}\ \bibnamefont {Hamann}}, \bibinfo {author} {\bibfnamefont {X.}~\bibnamefont {Gonze}}, \ and\ \bibinfo {author} {\bibfnamefont {G.-M.}\ \bibnamefont {Rignanese}},\ }\href@noop {} {\bibfield  {journal} {\bibinfo  {journal} {Computer Physics Communications}\ }\textbf {\bibinfo {volume} {226}},\ \bibinfo {pages} {39} (\bibinfo {year} {2018})}\BibitemShut {NoStop}%
\bibitem [{\citenamefont {Thompson}\ \emph {et~al.}(2022)\citenamefont {Thompson}, \citenamefont {Aktulga}, \citenamefont {Berger}, \citenamefont {Bolintineanu}, \citenamefont {Brown}, \citenamefont {Crozier}, \citenamefont {In't~Veld}, \citenamefont {Kohlmeyer}, \citenamefont {Moore}, \citenamefont {Nguyen} \emph {et~al.}}]{thompson2022lammps}%
  \BibitemOpen
  \bibfield  {author} {\bibinfo {author} {\bibfnamefont {A.~P.}\ \bibnamefont {Thompson}}, \bibinfo {author} {\bibfnamefont {H.~M.}\ \bibnamefont {Aktulga}}, \bibinfo {author} {\bibfnamefont {R.}~\bibnamefont {Berger}}, \bibinfo {author} {\bibfnamefont {D.~S.}\ \bibnamefont {Bolintineanu}}, \bibinfo {author} {\bibfnamefont {W.~M.}\ \bibnamefont {Brown}}, \bibinfo {author} {\bibfnamefont {P.~S.}\ \bibnamefont {Crozier}}, \bibinfo {author} {\bibfnamefont {P.~J.}\ \bibnamefont {In't~Veld}}, \bibinfo {author} {\bibfnamefont {A.}~\bibnamefont {Kohlmeyer}}, \bibinfo {author} {\bibfnamefont {S.~G.}\ \bibnamefont {Moore}}, \bibinfo {author} {\bibfnamefont {T.~D.}\ \bibnamefont {Nguyen}},  \emph {et~al.},\ }\href@noop {} {\bibfield  {journal} {\bibinfo  {journal} {Computer Physics Communications}\ }\textbf {\bibinfo {volume} {271}},\ \bibinfo {pages} {108171} (\bibinfo {year} {2022})}\BibitemShut {NoStop}%
\bibitem [{\citenamefont {Lehoucq}\ \emph {et~al.}(1998)\citenamefont {Lehoucq}, \citenamefont {Sorensen},\ and\ \citenamefont {Yang}}]{lehoucq1998arpack}%
  \BibitemOpen
  \bibfield  {author} {\bibinfo {author} {\bibfnamefont {R.~B.}\ \bibnamefont {Lehoucq}}, \bibinfo {author} {\bibfnamefont {D.~C.}\ \bibnamefont {Sorensen}}, \ and\ \bibinfo {author} {\bibfnamefont {C.}~\bibnamefont {Yang}},\ }\href@noop {} {\emph {\bibinfo {title} {{ARPACK} users' guide: solution of large-scale eigenvalue problems with implicitly restarted {Arnoldi} methods}}}\ (\bibinfo  {publisher} {SIAM},\ \bibinfo {year} {1998})\BibitemShut {NoStop}%
\bibitem [{\citenamefont {Virtanen}\ \emph {et~al.}(2020)\citenamefont {Virtanen}, \citenamefont {Gommers}, \citenamefont {Oliphant}, \citenamefont {Haberland}, \citenamefont {Reddy}, \citenamefont {Cournapeau}, \citenamefont {Burovski}, \citenamefont {Peterson}, \citenamefont {Weckesser}, \citenamefont {Bright}, \citenamefont {{van der Walt}}, \citenamefont {Brett}, \citenamefont {Wilson}, \citenamefont {Millman}, \citenamefont {Mayorov}, \citenamefont {Nelson}, \citenamefont {Jones}, \citenamefont {Kern}, \citenamefont {Larson}, \citenamefont {Carey}, \citenamefont {Polat}, \citenamefont {Feng}, \citenamefont {Moore}, \citenamefont {{VanderPlas}}, \citenamefont {Laxalde}, \citenamefont {Perktold}, \citenamefont {Cimrman}, \citenamefont {Henriksen}, \citenamefont {Quintero}, \citenamefont {Harris}, \citenamefont {Archibald}, \citenamefont {Ribeiro}, \citenamefont {Pedregosa}, \citenamefont {{van Mulbregt}},\ and\ \citenamefont {{SciPy 1.0 Contributors}}}]{2020SciPy-NMeth}%
  \BibitemOpen
  \bibfield  {author} {\bibinfo {author} {\bibfnamefont {P.}~\bibnamefont {Virtanen}}, \bibinfo {author} {\bibfnamefont {R.}~\bibnamefont {Gommers}}, \bibinfo {author} {\bibfnamefont {T.~E.}\ \bibnamefont {Oliphant}}, \bibinfo {author} {\bibfnamefont {M.}~\bibnamefont {Haberland}}, \bibinfo {author} {\bibfnamefont {T.}~\bibnamefont {Reddy}}, \bibinfo {author} {\bibfnamefont {D.}~\bibnamefont {Cournapeau}}, \bibinfo {author} {\bibfnamefont {E.}~\bibnamefont {Burovski}}, \bibinfo {author} {\bibfnamefont {P.}~\bibnamefont {Peterson}}, \bibinfo {author} {\bibfnamefont {W.}~\bibnamefont {Weckesser}}, \bibinfo {author} {\bibfnamefont {J.}~\bibnamefont {Bright}}, \bibinfo {author} {\bibfnamefont {S.~J.}\ \bibnamefont {{van der Walt}}}, \bibinfo {author} {\bibfnamefont {M.}~\bibnamefont {Brett}}, \bibinfo {author} {\bibfnamefont {J.}~\bibnamefont {Wilson}}, \bibinfo {author} {\bibfnamefont {K.~J.}\ \bibnamefont {Millman}}, \bibinfo {author} {\bibfnamefont {N.}~\bibnamefont {Mayorov}}, \bibinfo {author} {\bibfnamefont
  {A.~R.~J.}\ \bibnamefont {Nelson}}, \bibinfo {author} {\bibfnamefont {E.}~\bibnamefont {Jones}}, \bibinfo {author} {\bibfnamefont {R.}~\bibnamefont {Kern}}, \bibinfo {author} {\bibfnamefont {E.}~\bibnamefont {Larson}}, \bibinfo {author} {\bibfnamefont {C.~J.}\ \bibnamefont {Carey}}, \bibinfo {author} {\bibfnamefont {{\.I}.}~\bibnamefont {Polat}}, \bibinfo {author} {\bibfnamefont {Y.}~\bibnamefont {Feng}}, \bibinfo {author} {\bibfnamefont {E.~W.}\ \bibnamefont {Moore}}, \bibinfo {author} {\bibfnamefont {J.}~\bibnamefont {{VanderPlas}}}, \bibinfo {author} {\bibfnamefont {D.}~\bibnamefont {Laxalde}}, \bibinfo {author} {\bibfnamefont {J.}~\bibnamefont {Perktold}}, \bibinfo {author} {\bibfnamefont {R.}~\bibnamefont {Cimrman}}, \bibinfo {author} {\bibfnamefont {I.}~\bibnamefont {Henriksen}}, \bibinfo {author} {\bibfnamefont {E.~A.}\ \bibnamefont {Quintero}}, \bibinfo {author} {\bibfnamefont {C.~R.}\ \bibnamefont {Harris}}, \bibinfo {author} {\bibfnamefont {A.~M.}\ \bibnamefont {Archibald}}, \bibinfo {author}
  {\bibfnamefont {A.~H.}\ \bibnamefont {Ribeiro}}, \bibinfo {author} {\bibfnamefont {F.}~\bibnamefont {Pedregosa}}, \bibinfo {author} {\bibfnamefont {P.}~\bibnamefont {{van Mulbregt}}}, \ and\ \bibinfo {author} {\bibnamefont {{SciPy 1.0 Contributors}}},\ }\href {\doibase 10.1038/s41592-019-0686-2} {\bibfield  {journal} {\bibinfo  {journal} {Nature Methods}\ }\textbf {\bibinfo {volume} {17}},\ \bibinfo {pages} {261} (\bibinfo {year} {2020})}\BibitemShut {NoStop}%
\bibitem [{\citenamefont {Malenfant-Thuot}\ \emph {et~al.}(2024{\natexlab{b}})\citenamefont {Malenfant-Thuot}, \citenamefont {Shaaban~Kabakibo},\ and\ \citenamefont {Côté}}]{codes1}%
  \BibitemOpen
  \bibfield  {author} {\bibinfo {author} {\bibfnamefont {O.}~\bibnamefont {Malenfant-Thuot}}, \bibinfo {author} {\bibfnamefont {D.}~\bibnamefont {Shaaban~Kabakibo}}, \ and\ \bibinfo {author} {\bibfnamefont {M.}~\bibnamefont {Côté}},\ }\href@noop {} {\enquote {\bibinfo {title} {{ML Raman} repository},}\ }\bibinfo {howpublished} {\url{https://github.com/OMalenfantThuot/ml_raman}} (\bibinfo {year} {2024}{\natexlab{b}})\BibitemShut {NoStop}%
\bibitem [{\citenamefont {Malenfant-Thuot}\ \emph {et~al.}(2024{\natexlab{c}})\citenamefont {Malenfant-Thuot}, \citenamefont {Shaaban~Kabakibo},\ and\ \citenamefont {Côté}}]{codes2}%
  \BibitemOpen
  \bibfield  {author} {\bibinfo {author} {\bibfnamefont {O.}~\bibnamefont {Malenfant-Thuot}}, \bibinfo {author} {\bibfnamefont {D.}~\bibnamefont {Shaaban~Kabakibo}}, \ and\ \bibinfo {author} {\bibfnamefont {M.}~\bibnamefont {Côté}},\ }\href@noop {} {\enquote {\bibinfo {title} {{MlCalcDriver} repository},}\ }\bibinfo {howpublished} {\url{https://github.com/OMalenfantThuot/ML_Calc_Driver}} (\bibinfo {year} {2024}{\natexlab{c}})\BibitemShut {NoStop}%
\bibitem [{\citenamefont {Can{\c{c}}ado}\ \emph {et~al.}(2011)\citenamefont {Can{\c{c}}ado}, \citenamefont {Jorio}, \citenamefont {Ferreira}, \citenamefont {Stavale}, \citenamefont {Achete}, \citenamefont {Capaz}, \citenamefont {Moutinho}, \citenamefont {Lombardo}, \citenamefont {Kulmala},\ and\ \citenamefont {Ferrari}}]{canccado2011quantifying}%
  \BibitemOpen
  \bibfield  {author} {\bibinfo {author} {\bibfnamefont {L.~G.}\ \bibnamefont {Can{\c{c}}ado}}, \bibinfo {author} {\bibfnamefont {A.}~\bibnamefont {Jorio}}, \bibinfo {author} {\bibfnamefont {E.~M.}\ \bibnamefont {Ferreira}}, \bibinfo {author} {\bibfnamefont {F.}~\bibnamefont {Stavale}}, \bibinfo {author} {\bibfnamefont {C.~A.}\ \bibnamefont {Achete}}, \bibinfo {author} {\bibfnamefont {R.~B.}\ \bibnamefont {Capaz}}, \bibinfo {author} {\bibfnamefont {M.~d.~O.}\ \bibnamefont {Moutinho}}, \bibinfo {author} {\bibfnamefont {A.}~\bibnamefont {Lombardo}}, \bibinfo {author} {\bibfnamefont {T.}~\bibnamefont {Kulmala}}, \ and\ \bibinfo {author} {\bibfnamefont {A.~C.}\ \bibnamefont {Ferrari}},\ }\href@noop {} {\bibfield  {journal} {\bibinfo  {journal} {Nano letters}\ }\textbf {\bibinfo {volume} {11}},\ \bibinfo {pages} {3190} (\bibinfo {year} {2011})}\BibitemShut {NoStop}%
\bibitem [{\citenamefont {Wu}\ \emph {et~al.}(2018)\citenamefont {Wu}, \citenamefont {Lin}, \citenamefont {Cong}, \citenamefont {Liu},\ and\ \citenamefont {Tan}}]{wu2018raman}%
  \BibitemOpen
  \bibfield  {author} {\bibinfo {author} {\bibfnamefont {J.-B.}\ \bibnamefont {Wu}}, \bibinfo {author} {\bibfnamefont {M.-L.}\ \bibnamefont {Lin}}, \bibinfo {author} {\bibfnamefont {X.}~\bibnamefont {Cong}}, \bibinfo {author} {\bibfnamefont {H.-N.}\ \bibnamefont {Liu}}, \ and\ \bibinfo {author} {\bibfnamefont {P.-H.}\ \bibnamefont {Tan}},\ }\href@noop {} {\bibfield  {journal} {\bibinfo  {journal} {Chemical Society Reviews}\ }\textbf {\bibinfo {volume} {47}},\ \bibinfo {pages} {1822} (\bibinfo {year} {2018})}\BibitemShut {NoStop}%
\bibitem [{\citenamefont {Can{\c{c}}ado}\ \emph {et~al.}(2017)\citenamefont {Can{\c{c}}ado}, \citenamefont {Da~Silva}, \citenamefont {Ferreira}, \citenamefont {Hof}, \citenamefont {Kampioti}, \citenamefont {Huang}, \citenamefont {P{\'e}nicaud}, \citenamefont {Achete}, \citenamefont {Capaz},\ and\ \citenamefont {Jorio}}]{canccado2017disentangling}%
  \BibitemOpen
  \bibfield  {author} {\bibinfo {author} {\bibfnamefont {L.~G.}\ \bibnamefont {Can{\c{c}}ado}}, \bibinfo {author} {\bibfnamefont {M.~G.}\ \bibnamefont {Da~Silva}}, \bibinfo {author} {\bibfnamefont {E.~H.~M.}\ \bibnamefont {Ferreira}}, \bibinfo {author} {\bibfnamefont {F.}~\bibnamefont {Hof}}, \bibinfo {author} {\bibfnamefont {K.}~\bibnamefont {Kampioti}}, \bibinfo {author} {\bibfnamefont {K.}~\bibnamefont {Huang}}, \bibinfo {author} {\bibfnamefont {A.}~\bibnamefont {P{\'e}nicaud}}, \bibinfo {author} {\bibfnamefont {C.~A.}\ \bibnamefont {Achete}}, \bibinfo {author} {\bibfnamefont {R.~B.}\ \bibnamefont {Capaz}}, \ and\ \bibinfo {author} {\bibfnamefont {A.}~\bibnamefont {Jorio}},\ }\href@noop {} {\bibfield  {journal} {\bibinfo  {journal} {2D Materials}\ }\textbf {\bibinfo {volume} {4}},\ \bibinfo {pages} {025039} (\bibinfo {year} {2017})}\BibitemShut {NoStop}%
\bibitem [{\citenamefont {Bonini}\ \emph {et~al.}(2007)\citenamefont {Bonini}, \citenamefont {Lazzeri}, \citenamefont {Marzari},\ and\ \citenamefont {Mauri}}]{bonini2007phonon}%
  \BibitemOpen
  \bibfield  {author} {\bibinfo {author} {\bibfnamefont {N.}~\bibnamefont {Bonini}}, \bibinfo {author} {\bibfnamefont {M.}~\bibnamefont {Lazzeri}}, \bibinfo {author} {\bibfnamefont {N.}~\bibnamefont {Marzari}}, \ and\ \bibinfo {author} {\bibfnamefont {F.}~\bibnamefont {Mauri}},\ }\href@noop {} {\bibfield  {journal} {\bibinfo  {journal} {Physical review letters}\ }\textbf {\bibinfo {volume} {99}},\ \bibinfo {pages} {176802} (\bibinfo {year} {2007})}\BibitemShut {NoStop}%
\bibitem [{\citenamefont {Liu}\ \emph {et~al.}(2019)\citenamefont {Liu}, \citenamefont {Cong}, \citenamefont {Lin},\ and\ \citenamefont {Tan}}]{liu2019intrinsic}%
  \BibitemOpen
  \bibfield  {author} {\bibinfo {author} {\bibfnamefont {H.-N.}\ \bibnamefont {Liu}}, \bibinfo {author} {\bibfnamefont {X.}~\bibnamefont {Cong}}, \bibinfo {author} {\bibfnamefont {M.-L.}\ \bibnamefont {Lin}}, \ and\ \bibinfo {author} {\bibfnamefont {P.-H.}\ \bibnamefont {Tan}},\ }\href@noop {} {\bibfield  {journal} {\bibinfo  {journal} {Carbon}\ }\textbf {\bibinfo {volume} {152}},\ \bibinfo {pages} {451} (\bibinfo {year} {2019})}\BibitemShut {NoStop}%
\bibitem [{\citenamefont {Adamyan}\ and\ \citenamefont {Zavalniuk}(2010)}]{adamyan2010phonons}%
  \BibitemOpen
  \bibfield  {author} {\bibinfo {author} {\bibfnamefont {V.}~\bibnamefont {Adamyan}}\ and\ \bibinfo {author} {\bibfnamefont {V.}~\bibnamefont {Zavalniuk}},\ }\href@noop {} {\bibfield  {journal} {\bibinfo  {journal} {Journal of Physics: Condensed Matter}\ }\textbf {\bibinfo {volume} {23}},\ \bibinfo {pages} {015402} (\bibinfo {year} {2010})}\BibitemShut {NoStop}%
\bibitem [{\citenamefont {Koukaras}\ \emph {et~al.}(2015)\citenamefont {Koukaras}, \citenamefont {Kalosakas}, \citenamefont {Galiotis},\ and\ \citenamefont {Papagelis}}]{koukaras2015phonon}%
  \BibitemOpen
  \bibfield  {author} {\bibinfo {author} {\bibfnamefont {E.~N.}\ \bibnamefont {Koukaras}}, \bibinfo {author} {\bibfnamefont {G.}~\bibnamefont {Kalosakas}}, \bibinfo {author} {\bibfnamefont {C.}~\bibnamefont {Galiotis}}, \ and\ \bibinfo {author} {\bibfnamefont {K.}~\bibnamefont {Papagelis}},\ }\href@noop {} {\bibfield  {journal} {\bibinfo  {journal} {Scientific reports}\ }\textbf {\bibinfo {volume} {5}},\ \bibinfo {pages} {12923} (\bibinfo {year} {2015})}\BibitemShut {NoStop}%
\end{thebibliography}%

\end{document}